\documentclass[a4paper,12pt]{article}
\usepackage{amssymb, amsmath}
\usepackage{epsfig}
\makeatletter
\renewcommand{\section}{{\setcounter{equation}{0}}\@startsection%
{section}%
{1}%
{0mm}%
{-\baselineskip}%
{0.5\baselineskip}%
{\normalfont\large\bfseries}%
} \makeatother
\renewcommand{\theequation}{\arabic{section}.\arabic{equation}}

\makeatletter
\renewcommand{\subsection}{\@startsection%
{subsection}%
{2}%
{0mm}%
{-\baselineskip}%
{0.5\baselineskip}%
{\normalfont\normalsize\bfseries}}%

\newtheorem{remark}{Remark}[section]
\newtheorem{proposition}{Proposition}[section]
\newtheorem{theorem}{Theorem}[section]

\newtheorem{definition}{Definition}

\newenvironment{proof}{{\bf Proof:}}{\hfill$\square$\vskip.5cm}

\usepackage{color}

\def\*{{\phantom *}}

\newcommand{\sV}{{\hbox {\tiny{$V$}}}}
\newcommand{\sbeta}{{\hbox {\tiny{$\beta$}}}}

\newcommand{\Hone}{{\mathcal{H}_\sV}}

\newcommand{\trace}{\text{\rm{trace}}\,}

\newcommand{\Prob}{\mathbb{P}}
\newcommand{\Ex}{\mathbb{E}}

\newcommand{\starr}{^{\phantom *}}

\newcommand{\cHam}{H_\sV^{(n)}}   
\newcommand{\hcHam}{H^{\mathsf{hc}}_{n,\sV}}      
\newcommand{\cEx}{\Ex^n_\sV}   
\newcommand{\cProb}{\Prob^n_\sV}   
\newcommand{\cPart}{Z_\sbeta(n,V)}           
\newcommand{\ccPart}{Z_\sbeta\left({(V\!\!\!-\!q)}/{V},\, n\!-\!q,V\!\!\!-\!q\right)}

\newcommand{\CC}{\mathbb{C}}
\newcommand{\RR}{\mathbb{R}}

\newcommand{\la}{\langle}
\newcommand{\ra}{\rangle}

\newcommand{\Phc}{\mathcal{P}_n^{\mathsf{hc}}}
\newcommand{\Hcansym}[1]{{\mathcal{H}_{\sV,+}^{(#1)}}}
\newcommand{\Hcan}[1]{{\mathcal{H}_{\sV}^{(#1)}}}
\newcommand{\Hhccan}[1]{{\mathcal{H}_{#1,\sV}^{\mathsf{hc}}}}
\newcommand{\Hhccansym}[1]{{\mathcal{H}_{#1,\sV,+}^{{\mathsf{hc}}}}}
\newcommand{\Hhccyclespace}{{\mathcal{H}_{q,n,\sV}^{{\mathsf{hc}}}}}

\newcommand{\Htilde}{\widetilde{H}}
\newcommand{\ii}{\mathbf{i}}
\newcommand{\jj}{\mathbf{j}}
\newcommand{\kk}{\mathbf{k}}
\newcommand{\boldl}{\mathbf{l}}
\newcommand{\e}{\mathrm{e}}

\newcommand{\thermlim}{ \lim_{\stackrel{n,\sV \to \infty}{n/\sV=\rho}}}

\begin{document}


\phantom{.} 
\textbf{To appear in JSP}\hfill
\textbf{04-06-08}
\vskip1.5cm
\begin{center}
{\Large Long Cycles
\vskip 0.1cm
in the Infinite-Range-Hopping Bose-Hubbard Model
\vskip 0.1cm
with Hard Cores}
\vskip 0.5cm
{\bf G. Boland}
\footnote{email: Gerry.Boland@ucd.ie}
\vskip 0.1cm
and
\vskip 0.1cm
{\bf
J.V. Pul\'e}
\footnote{email:
Joe.Pule@ucd.ie}
\vskip 0.5cm
School of Mathematical Sciences
\linebreak
University College Dublin\\Belfield, Dublin 4, Ireland
\end{center}
\vskip 1cm
\begin{abstract}
\vskip -0.4truecm
\noindent
In this paper we study the relation between long cycles and Bose-Condensation in the Infinite range Bose-Hubbard Model with a hard core interaction.
We calculate the density of particles on long cycles in the thermodynamic limit
and find that the existence
of a non-zero long cycle density coincides with the occurrence of Bose-Einstein
condensation but this density is not equal to that of the Bose condensate.

\end{abstract}
\newpage\setcounter{page}{1}
\section{Introduction}
In 1953, Feynman analysed the partition function of an interacting Bose gas in terms
of the statistical distribution
of permutation cycles of particles and emphasized the roles of long cycles at the
transition point \cite{Fey 1953 and stat.mech.}. Then Penrose and Onsager,
pursuing Feynman's arguments,
observed that there should be Bose condensation when the fraction of the total particle
number belonging to
long cycles is strictly positive \cite{Pen.Onsa.}.
These ideas are now commonly accepted and also discussed in various contexts
in systems showing analogous phase
transitions such as percolation, gelation and polymerization
(see e.g. \cite{Chandler}, \cite{Sear-Cuesta}, \cite{Schakel}), though
it has been recently argued by Ueltschi \cite{Ueltschi1}
that in fact the hypothesis is not always valid. To our knowledge, there had not
appeared a precise mathematical and quantitative formulation of the relation
between Bose condensate and long cycles until the work of S\"ut\"o \cite{Suto} and
its validity
has been checked only in a few models: the free and mean field Bose gas in \cite{Suto},
(see also Ueltschi \cite{Ueltschi2})
and the perturbed mean field model of a Bose gas studied in \cite{DMP}.
In these models it is shown that the density
of particles in long cycles is equal to the Bose condensate density.
The purpose of this paper is test the validity of the hypothesis in yet another
model of a Bose gas, the Infinite range Bose-Hubbard Model with a hard core.
Here we calculate the density of particles on long cycles in the thermodynamic limit
and find that though the existence
of a non-zero long cycle density coincides with the occurence of Bose-Einstein
condensation, this density is not equal to the Bose condensate density.
\par
The main simplifying feature in this model is the following. In general the density
of particles on cycles
of length $q$ for $n$ particles can be expressed (apart from normalization)
as the trace (see for example Proposition \ref{stat})
of the exponential of the Hamiltonian for $n-q$ bosons and $q$ distinguishable
particles (no statistics). In terms of the random walk representation (cf \cite{Toth}), the particles in this model are allowed to hop from one site
to another with equal probability. We can prove (Proposition \ref{c}) that in the thermodynamic limit
we can neglect the hopping of the $q$ particles so that bosons have to avoid
each other and the fixed positions of the
distinguishable particles. This is equivalent to a reduction of the
lattice by $q$ sites. Moreover the $q$ particles are on a cycle of length $q$.
For $q > 1$, this means for example, that the position of the second particle at the beginning of its path is same as the position of the first particle at the end of its path. But since they do not hop this is impossible by the hard core condition
and therefore among the short cycles only the cycle of unit length contributes. Since
the sum of all the cycle densities gives the particle density, this means
that in the thermodynamic limit
the sum of the long cycle densities is the particle density less the one-cycle contribution.
The one-cycle density, apart from some scaling and the normalization, is then the partition
function for the boson system with one site removed from the lattice, which can be calculated.
\par
The model without a hard-core will be treated in another paper. There we can again neglect the hopping of the  $q$ distinguishable
particles. However in that case cycles of all lengths contribute to the long-cycle density. It is relatively easy to see that when there is no condensation
the long-cycle density vanishes but we do not yet know what happens when there is Bose-Einstein condensation.
\par
In Section 2 we first describe the model and recall its thermodynamic properties
as stated by Penrose \cite{Penrose} (see also T\'oth \cite{Toth} and Kirson\cite{Kirson}).
We then apply the general framework for cycle statistics described in \cite{DMP}, following
\cite{Martin}.
Using standard properties of the decomposition of permutations into cycles,
the canonical sum is
converted into a sum on cycle lengths.
This makes it possible to decompose the total density $\rho=\rho_{{\rm short}}+\rho_{{\rm long}}$
into the number density of particles belonging to cycles of finite length
($\rho_{{\rm short}}$)
and to infinitely long cycles ($\rho_{{\rm long}}$) in the thermodynamic limit.
It is conjectured
that when there is Bose condensation, $\rho_{{\rm long}}$ is different from
zero and identical to the condensate density. The main purpose of the paper is to check the
validity of this conjecture in our model. At the end of Section 2 we state in the main theorem
describing the relation between Bose-Einstein condensation and the density of long
cycles for our model.
\par
In Section 3 we prove the main theorem and in Section 4 we discuss briefly \textit{Off-diagonal
Long-Range Order}.
\section{The Model and Results}
The Bose-Hubbard Hamiltonian is given by
\begin{equation}\label{B-H}
    H^{\mathrm{BH}}=J\!\!\!\!\!\!\sum_{x,y\in\Lambda_\sV\,:|x-y|=1}(a^*_x-a^*_y)(a\starr_x-a\starr_y)
    +\lambda\sum_{x\in \Lambda_{\sV}} n_x(n_x-1)
\end{equation}
where $\Lambda_\sV$ is a lattice of $V$ sites, $a^*_x$ and $a\starr_x$ are the Bose
creation and annihilation operators satisfying the usual commutation relations
$[a^*_x,a\starr_y]=\delta_{x,y}$ and $n_x=a^*_xa^{\phantom *}_x$.
The first term with $J>0$ is the kinetic energy operator and the
second term with $\lambda>0$ describes a repulsive interaction, as it discourages the presence of more
than one particle at each site. This model was originally introduced by
Fisher \textit{et al.} \cite{Fisher}.

The infinite-range hopping model is given by the Hamiltonian
\begin{equation}\label{I-R}
    H^{\mathrm{IR}}=\frac{1}{2V}\!\!\!\sum_{x,y\in\Lambda_\sV}(a^*_x-a^*_y)(a\starr_x-a\starr_y)
    +\lambda\sum_{x\in \Lambda_\sV} n_x(n_x-1).
\end{equation}
This is in fact a mean-field version of (\ref{B-H}) but in terms of the kinetic energy rather
than the interaction. In particular as in all mean-field models, the
lattice structure is irrelevant and there is no dependence
on dimensionality, so we can take $\Lambda_\sV=\{1,2,3, \ldots, V\}$.
The non-zero temperature properties of this model have
been studied by Bru and Dorlas \cite{BruDorlas} and by Adams and Dorlas \cite{AdamsDorlas}. We shall study
a special case of (\ref{I-R}), introduced by T\'oth \cite{Toth} where $\lambda=+\infty$,
that is complete single-site exclusion (hard-core). The properties of this model
in the canonical ensemble were first obtained by T\'oth using probabilistic methods. Later
Penrose \cite{Penrose} and Kirson \cite{Kirson} obtained equivalent results.
In the grand-canonical
ensemble the model is equivalent to the strong-coupling BCS model
(see for example Angelescu \cite{Angelescu}). Here we recall the thermodynamic properties of the
model in the canonical ensemble as given by Penrose.
\par
\begin{figure}[hbt]
\begin{center}
\includegraphics[width=10cm]{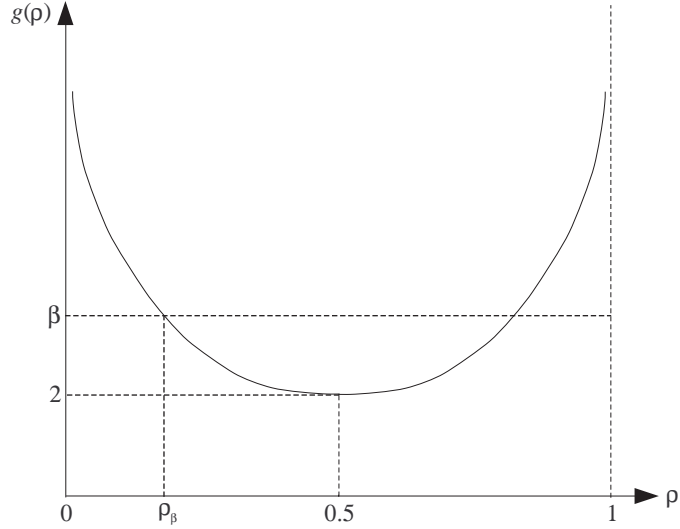}
\end{center}
\caption{\it Definition of $\rho_\sbeta$}
\label{fig1}
\end{figure}

For $\rho\in(0, 1)$, let
\begin{equation*}\label{g}
    g(\rho)=
    \begin{cases}
    {\displaystyle  \frac{1}{1-2\rho}\ln\left( \frac{1-\rho}{\rho}\right )} \vspace{0.2cm}
    & \mathrm{if}\ \  \rho\neq 1/2,\\
    2 & \mathrm{if}\ \ \rho=1/2.
    \end{cases}
\end{equation*}
For each $\beta\geq 2$ the equation $\beta=g(\rho)$ has a unique solution in $(0,1/2]$ denoted by
$\rho_\beta$ (see Fig.\ref{fig1}). We define $\rho_\beta:=1/2$ for $\beta<2 $.
\begin{theorem} $\mathrm{(Penrose\ [10],\ Theorem\ 1)}$\\
The free energy per site at inverse temperature $\beta$ as a function of the particle density
$\rho\in[0, 1]$, $f_\beta(\rho)$, is given by
\begin{equation*}\label{f}
    f_\beta(\rho)=
    \begin{cases}
    {\displaystyle \rho+\frac{1}{\beta}\left(\rho\ln\rho+(1-\rho)\ln(1-\rho)\right )}
    & \mathrm{if}\ \  \rho\in[0,\rho_\beta]\cup[1-\rho_\beta,1],\\
    {\displaystyle \rho^2 +\rho_\beta(1-\rho_\beta)+\frac{1}{\beta}
    \left(\rho_\beta\ln\rho_\beta+(1-\rho_\beta)\ln(1-\rho_\beta)\right )}
    & \mathrm{if}\ \ \rho\in[\rho_\beta,1-\rho_\beta].
    \end{cases}
\end{equation*}
\end{theorem}
The density of particles in the ground state in the thermodynamic limit is given by
\begin{equation}\label{}
    \rho_\beta^c=\thermlim\frac{1}{V^2}\hskip -0.2cm\sum_{x,y\in\Lambda_\sV}\la a^*_x a\starr_y\ra
\end{equation}
where $\la\,\cdot\,\ra$ denotes the canonical expectation for $n$ particles.
Penrose showed that for certain values of $\rho$ and $\beta$, Bose-Einstein condensation occurs,
that is, $\rho_\beta^c>0$. The Bose-condensate density is given in the following theorem.
\begin{theorem} $\mathrm{(Penrose\ [10],\ Theorem\ 2)}$\\
The Bose-condensate density, $\rho_\beta^c$ at inverse temperature $\beta$ as a function
of the particle density
$\rho\in[0, 1]$, is given by
\begin{equation*}\label{rho-c}
	\rho_\beta^c=
	\begin{cases}
		0
		& \mathrm{if}\ \  \rho\in[0,\rho_\beta]\cup[1-\rho_\beta,1],\\
		{\displaystyle (\rho-\rho_\beta)(1-\rho-\rho_\beta)}
		& \mathrm{if}\ \ \rho\in[\rho_\beta,1-\rho_\beta].
    	\end{cases}
\end{equation*}
\end{theorem}
We note that both $f_\beta(\rho)-\rho$ and the condensate density $\rho_\beta^c$ are
symmetric about $\rho=1/2$.
This can easily seen by interchanging particles and holes. The Boson states being symmetric can be labelled
unambiguously by the sites they occupy but equivalently they can be labelled by the sites they do not occupy (holes).
\par
Before proceeding to the study of cycle statistics we need to define the $n$-particle
Hamiltonian more carefully. The single particle Hilbert space is $\mathcal{H}_\sV:=\CC^\sV$
and on it we define the operator
\begin{equation*}
\label{ham1}
	H_\sV=I-P_\sV
\end{equation*}
where $P_\sV$ is the orthogonal projection onto the unit vector
\begin{equation*}
	\mathbf{g}_\sV = \frac{1}{\sqrt{V}} (1,1,\dots,1)\in \mathcal{H}_\sV.
\end{equation*}
In terms of the usual basis vectors of $\mathcal{H}_\sV$, $\{\mathbf{e}_i\, |\, i=1\ldots V\}$, $P_\sV$ is given by
\begin{equation*}
	P_\sV  \mathbf{e}_i  = \frac{1}{V} \sum_{j=1}^V  \mathbf{e}_j.
\end{equation*}
Thus $H_\sV$ is the orthogonal projection onto the subspace orthogonal to $\mathbf{g}_\sV$.
For an operator $A$ on $\Hone$, we define $A^{(n)}$ on
$\Hcan{n} = \underbrace{\Hone \otimes \Hone \otimes \dots
\otimes \Hone}_{n \text{ times}}$ by \vspace{-0.2cm}
\begin{equation*}\label{A-n}
	A^{(n)}
	=A\otimes I\otimes \ldots \otimes I+I\otimes A\otimes \ldots \otimes I+\ldots
		+I\otimes I\otimes \ldots \otimes A.
\end{equation*}
With this notation we can define the non-interacting $n$-particle Hamiltonian $\cHam$ acting
on the unsymmetrised Hilbert space $\Hcan{n}$ as:
\begin{align*}
	\cHam
	&= I^{(n)} - P_\sV^{(n)}
\\
 	&= n - P_\sV \otimes I \otimes \dots \otimes I - I \otimes P_\sV \otimes I \otimes \dots \otimes I
                	- \dots - I \otimes I \otimes \dots \otimes P_\sV.
\end{align*}
For bosons we have to consider the symmetric subspace of $\Hcan{n}$.
The symmetrisation projection $\sigma_{+}^n$ on $\Hcan{n}$ is defined by
\begin{equation}\label{sym}
	\sigma_{+}^n = \frac{1}{n!} \sum_{\pi \in S_n} U_\pi
\end{equation}
where $U_\pi: \Hcan{n} \mapsto \Hcan{n}$ is the unitary representation of the permutation group
$S_n$ on $\Hcan{n}$ defined by
\begin{equation*}
	U_\pi(\phi_1 \otimes \phi_2 \otimes \cdots \otimes \phi_n)
	= \phi_{\pi(1)} \otimes \phi_{\pi(2)} \otimes \cdots
		\otimes \phi_{\pi(n)}, \;\; \phi_j \in \Hone, \; j=1, \dots, n; \;  \pi \in S_n.
\end{equation*}
Then the symmetric $n$-particle subspace is $\Hcansym{n} := \sigma_{+}^n \Hcan{n}$.
\par
When $\cHam$ is restricted to $\Hcansym{n}$, we obtain
\begin{equation*}
	\frac{1}{2V}\!\!\!\sum_{x,y\in\Lambda_\sV}(a^*_x-a^*_y)(a\starr_x-a\starr_y).
\end{equation*}
We introduce the hard-core interaction by applying a projection
to $\Hcan{n}$ to forbid more than one particle from occupying each site.
Let $\{\mathbf{e}_i\}_{i=1}^\sV$ be the usual orthonormal basis for $\Hone$.
We then define the hard core projection $\Phc$
on $\Hcan{n}$ by
\begin{equation}\label{P-hardcore}
	\Phc ( \mathbf{e}_{i_1} \otimes \mathbf{e}_{i_2} \otimes \dots \otimes \mathbf{e}_{i_n} ) =
	\begin{cases}
 		0   & \text{if } \, \mathbf{e}_{i_k} = \mathbf{e}_{i_{k'}} \; \text{for some }\, k \ne k', \\
		\mathbf{e}_{i_1} \otimes \mathbf{e}_{i_2} \otimes \dots \otimes \mathbf{e}_{i_n}   & \text{otherwise}.
	\end{cases}
\end{equation}
We shall call $\Hhccan{n} := \Phc \Hcan{n}$  the unsymmetrised hard-core $n$-particle space
and \linebreak $\Hhccansym{n} := \Phc \Hcansym{n} $ the symmetric hard-core $n$-particle space.
Note that as $[U_\pi,\Phc ]=0$ for all $\pi\in S_n$, $\Phc$ commutes with the symmetrisation
and so $\Hhccansym{n} = \sigma_{+}^n \Hhccan{n}$.

The hard-core $n$-particle Hamiltonian is then
\begin{equation}\label{H-hc}
    \hcHam:=\Phc \cHam \Phc
\end{equation}
acting on the hard-core $n$-particle space $\Hhccan{n}$. Therefore the Hamiltonian for the
infinite-range Bose-Hubbard model with hard-core is (\ref{H-hc}) acting on the \textbf{symmetric} hard-core
$n$-particle space $\Hhccansym{n}$.
\par
We shall now analyse the cycle statistics of this model.
\par
Using (\ref{sym}), the canonical partition function for the hard-core boson model may be written as
\begin{equation*}\label{}
    \cPart = \trace_{\Hhccansym{n}} \left[ \e^{-\beta \hcHam} \right]
   = \trace_{\Hhccan{n}} \left[\sigma_{+}^n \e^{-\beta \hcHam} \right]
   = \frac{1}{n!} \sum_{\pi \in S_n} \trace_{\Hhccan{n}} \left[ U_\pi \e^{-\beta \hcHam} \right].
\end{equation*}
Following \cite{DMP}, we define a probability measure on the permutation group $S_n$ by
\begin{equation}
\label{cProb_defn} 
  \cProb(\pi) = \frac{1}{\cPart} \frac{1}{n!}
  \trace_{\Hhccan{n}} \left[ U_\pi \e^{-\beta \hcHam} \right].
\end{equation}
From the random walk formulation (see for example \cite{Toth}) one can see that the kernel of $\e^{-\beta \hcHam}$
is positive and therefore the righthand side of (\ref{cProb_defn}) is positive.
\par
\begin{figure}[hbt]
\begin{center}
\includegraphics[width=12cm]{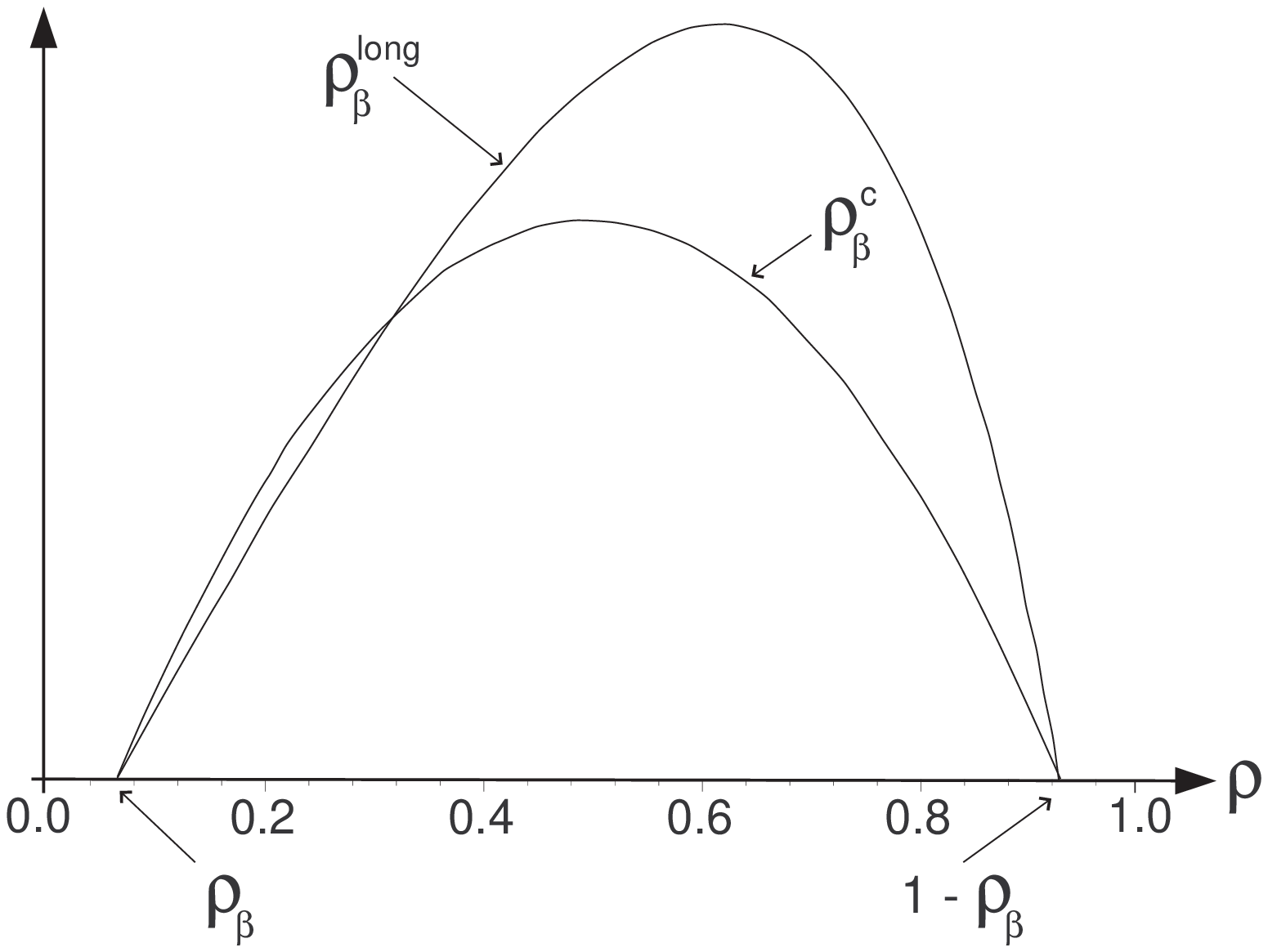}
\end{center}
\caption{\it $\rho_\sbeta^c$ and $\rho_\sbeta^{\mathrm{long}}$ for $\beta>2$}
\label{fig2}
\end{figure}

Each permutation $\pi \in S_n$ can be decomposed uniquely into a number of cyclic permutations of lengths
$q_1, q_2, \dots, q_r$ with $r \le n$ and $q_1 + q_2 + \dots + q_r = n$.
For $q \in \{1,2,\ldots,n\}$, let $N_q(\pi)$ be the random variable
corresponding to the number of cycles of length $q$ in $\pi$. Then
the expectation of the number of $q$-cycles in the canonical ensemble is:
\begin{equation*}\label{}
    \cEx(N_q) = \sum_{r=1}^n r\cProb(N_q\!=\!r)
\end{equation*}
and the average density of particles in $q$-cycles for the system of $n$ bosons is
\begin{equation}\label{}
    c_\sV^n(q) = \frac{q \; \cEx(N_q) }{V}.
\end{equation}
This brings us then to the following definition.
\begin{definition}
The expected density of particles on cycles of \textbf{infinite} length is given by
\begin{equation}\label{}
    \rho^{\mathrm{long}}_\sbeta = \lim_{Q \to \infty} \;\thermlim \; \sum_{q=Q+1}^n c_\sV^n(q).
\end{equation}
\end{definition}
For the free Bose gas, the mean field and the perturbed mean field Bose gas, it has been
shown that $\rho^{\mathrm{long}}_\sbeta=\rho_\sbeta^c$, the condensate density. For our model,
the situation is different. Below we state the main result of this paper:

\begin{theorem}\label{Th-rho-long}
The expected density of particles on cycles of infinite length,
$\rho_\sbeta^{\mathrm{long}}$, at inverse temperature $\beta$ as a function of the particle density
$\rho\in[0, 1]$, is given by
\begin{equation*}\label{rho-long}
   \rho_\sbeta^{\mathrm{long}}=
    \begin{cases}
    0
    & \mathrm{if}\ \  \rho\in[0,\rho_\sbeta]\cup[1-\rho_\sbeta,1],\\
    {\displaystyle \rho-\rho_\sbeta\mathrm{e}^{\beta(\rho-\rho_\sbeta)}}
    & \mathrm{if}\ \ \rho\in[\rho_\sbeta,1-\rho_\sbeta].
    \end{cases}
\end{equation*}
\end{theorem}
We note that (see Fig.\ref{fig2}):
\begin{itemize}
    \item $\rho_\sbeta^{\mathrm{long}}=0$ if and only if $\rho_\sbeta^c=0$.
    \item $\rho_\sbeta^{\mathrm{long}}$ is not symmetric with respect to $\rho=1/2$. As mentioned above the symmetry of the model about
    $\rho=1/2$ is due to the particle-hole symmetry. But the equivalent labelling of states by sets of occupied or unoccupied
    sites (particles and holes) cannot be used for distinguishable particles. We shall
    see (Proposition \ref{stat}) that the
    $q$-cycle occupation density $c_\sV^n(q)$ involves $q$ distinguishable particles and $n-q$ bosons and
    therefore the particle-hole symmetry is broken.
    \item When $\rho_\sbeta^c>0$, $\rho_\sbeta^{\mathrm{long}}$ starts below $\rho_\sbeta^c$ since its
    slope at $\rho_\beta$ is equal to $1-2\rho_\sbeta$ while $\rho_\sbeta^c$ has slope $1-\beta\rho_\sbeta$
    and $\beta>2$. Conversely, $\rho_\sbeta^{\mathrm{long}}$ finishes above
    $\rho_\sbeta^c$ since its slope at $1-\rho_\sbeta$ is less than that of $\rho_\sbeta^c$.
    \end{itemize}
%
%
%
%
%
\section{Proof of the Main Result}
In this section we shall prove Theorem \ref{Th-rho-long}. First we note that if $n/V=\rho$, then
\begin{equation}\label{n}
    \sum_{q=1}^n c^n_\sV(q)=\rho,
\end{equation}
so that if we define
\begin{equation*}\label{}
    \rho^{\mathrm{short}}_\sbeta = \lim_{Q \to \infty} \;\thermlim \; \sum_{q=1}^Q c_\sV^n(q)
\end{equation*}
we have
\begin{equation*}\label{}
    \rho^{\mathrm{long}}_\sbeta=\rho -\rho^{\mathrm{short}}_\sbeta.
\end{equation*}

For $\rho^{\mathrm{short}}_\sbeta$ we can interchange the sum over $q$ and $\thermlim$,
\begin{equation*}\label{}
    \rho^{\mathrm{short}}_\sbeta = \lim_{Q \to \infty} \; \sum_{q=1}^Q \thermlim \;c_\sV^n(q),
\end{equation*}
making it much easier to calculate. In fact we shall prove that:
\begin{equation}\label{rho-short}
   \rho_\sbeta^{\mathrm{short}}=
    \begin{cases}
    \rho
    & \mathrm{if}\ \  \rho\in[0,\rho_\sbeta]\cup[1-\rho_\sbeta,1],\\
    {\displaystyle \rho_\sbeta\mathrm{e}^{\beta(\rho-\rho_\sbeta)}}
    & \mathrm{if}\ \ \rho\in[\rho_\sbeta,1-\rho_\sbeta].
    \end{cases}
\end{equation}

The proof is in four steps. The first step is to obtain a convenient expression for
$c^n_\sV(q)$, the mean density of particles belonging to a cycle of length $q$.
We shall denote the unitary representation of a $q$-cycle by $U_q: \Hcan{q} \to \Hcan{q}$, that is
\begin{equation*}
	U_q ( \phi_{i_1} \otimes \dots \otimes \phi_{i_n}) =
	\phi_{i_2} \otimes \dots \phi_{i_q} \otimes \phi_{i_1} .
\end{equation*}
When there is no ambiguity we shall use the same notation $U_q$ for $U_q \otimes I^{(n-q)}: \Hcan{n} \to \Hcan{n}$
where $I$ is the identity operator. Note that $[U_q, \Phc] = 0$ and $[U_q, \sigma^n_{+}] = 0$.

\begin{proposition}\label{stat}
\begin{equation*}
	c^n_\sV(q) = \frac{1}{Z_\sbeta(n,V)} \frac{1}{V} \, \trace_{ \Hhccyclespace  }
				\left[ U_q \e^{-\beta H^{\mathsf{hc}}_{n,\sV} } \right]
\end{equation*}
where $\Hhccyclespace := \Phc ( \Hcan{q} \otimes \Hcansym{n-q} )$.
\end{proposition}
Note that though we write this proposition for our special case, in fact $c^n_\sV(q)$ can be expressed in this form
for any Boson model with a symmetric Hamiltonian.

By using cycle statistics, we split our symmetric hard-core Hilbert space $\Hhccansym{n}$
into a tensor product of two spaces, an unsymmetrised $q$-particle space $\Hcan{q}$ and a
symmetric $n-q$ particle space $\Hcansym{n-q}$, with the hard-core projection applied.
Writing
\begin{equation*}
    A^{(q)}:= A^{(q)} \otimes I^{(n-q)} \ \ \ \ \mathrm{and}
    \ \ \ \ A^{(n-q)} := I^{(q)} \otimes A^{(n-q)}
\end{equation*}
for any operator $A$ on $\mathcal{H}_\sV$,
we can express our Hamiltonian (\ref{H-hc}) on $\Hhccyclespace$ as follows:
\begin{equation*}\label{}
H^{\mathsf{hc}}_{n,\sV} = \Phc \left( n - P_\sV^{(q)} - P_\sV^{(n-q)} \right) \Phc.
\end{equation*}
Let $\widetilde{P}^{(q)}_\sV = \Phc P^{(q)}_\sV \Phc$ and
define the following reduced Hamiltonian
\begin{equation}\label{redHam}
   \widetilde{H}^{\mathsf{hc}}_{q,n,\sV}
   = \Phc \left(n - P_\sV^{(n-q)} \right) \Phc,
\end{equation}
so that
\begin{equation*}\label{}
    H^{\mathsf{hc}}_{n,\sV}
    = \widetilde{H}^{\mathsf{hc}}_{q,n,\sV} - \widetilde{P}^{(q)}_\sV.
\end{equation*}
The next step is to estimate the effect of neglecting the
action of the $\widetilde{P}^{(q)}_\sV$ term (equivalent to the hopping of the $q$ particles)
in the unsymmetrised space.
Let
\begin{equation}\label{c_tilde}
   \widetilde{c}\,^n_\sV(q) =  \frac{1}{Z_\sbeta(n,V)} \frac{1}{V} \trace_{\Hhccyclespace}
        \left[ U_q \mathrm{e}^{-\beta \widetilde{H}^{\mathsf{hc}}_{q,n,\sV} } \right],
\end{equation}
and define
\begin{equation*}\label{}
    Z_\sbeta(\lambda,n,V) = \trace_{\Hhccansym{n}}
    \left[ \mathrm{e}^{-\beta H^{\mathsf{hc}}_{\lambda, n,\sV} } \right]
\end{equation*}
where
\begin{equation}\label{h-lambda}
H^{\mathsf{hc}}_{\lambda, n,\sV} = \Phc \left( n - \lambda P_\sV^{(n)} \right) \Phc .
\end{equation}
Then we have the following estimate.
\begin{proposition}\label{c}
\begin{equation*}
    \left| c\,^n_\sV(q)-\widetilde{c}\,^n_\sV(q) \right|
\leq \frac{(1-\mathrm{e}^{-\beta q})}{V}
\frac{Z_\sbeta(\frac{V-q}{V},n-q,V-q)}{Z_\sbeta(n,V)}.
\end{equation*}
\end{proposition}
In the third step we obtain the limit of the ratio on the righthand side of the last inequality:
\begin{proposition}\label{Z}
\begin{equation}
   \thermlim\frac{Z_\sbeta(\frac{V-q}{V},n-q,V-q)}{Z_\sbeta(n,V)}=
    \begin{cases}
    \rho^q \, \e^{\beta q}
    & \mathrm{if}\ \  \rho\in[0,\rho_\sbeta]\cup[1-\rho_\sbeta,1],\\
    {\displaystyle \rho^q_\sbeta \, \mathrm{e}^{\beta q(1+\rho-\rho_\sbeta)}}
    & \mathrm{if}\ \ \rho\in[\rho_\sbeta,1-\rho_\sbeta].
    \end{cases}
\end{equation}
\end{proposition}
The final step is a simple proposition where we check the following:
\begin{proposition}\label{c-q}
$\widetilde{c}\,^n_\sV(q)=0$ if $q>1$ and
\begin{equation}
   \thermlim\widetilde{c}\,^n_\sV(1)=
    \begin{cases}
    \rho
    & \mathrm{if}\ \  \rho\in[0,\rho_\sbeta]\cup[1-\rho_\sbeta,1],\\
    {\displaystyle \rho_\sbeta \, \mathrm{e}^{\beta(\rho-\rho_\sbeta)}}
    & \mathrm{if}\ \ \rho\in[\rho_\sbeta,1-\rho_\sbeta].
    \end{cases}
\end{equation}
\end{proposition}
Using these four results the main result, Theorem \ref{Th-rho-long} follows very easily.
From Propositions \ref{c} and \ref{Z}
we have
\begin{equation*}
    \thermlim c\,^n_\sV(q)=\thermlim\widetilde{c}\,^n_\sV(q).
\end{equation*}
Since by Proposition \ref{c-q}, $\widetilde{c}\,^n_\sV(q)$=0 if $q>1$, it follows that
\begin{equation*}\label{rho-short-2}
   \rho_\sbeta^{\mathrm{short}}=\thermlim c\,^n_\sV(1)=
    \begin{cases}
    \rho
    & \mathrm{if}\ \  \rho\in[0,\rho_\sbeta]\cup[1-\rho_\sbeta,1],\\
    {\displaystyle \rho_\sbeta\mathrm{e}^{\beta(\rho-\rho_\sbeta)}}
    & \mathrm{if}\ \ \rho\in[\rho_\sbeta,1-\rho_\sbeta]
    \end{cases}
\end{equation*}
which is the required result.
\par
In the next four subsections we prove the results stated above.
%
\subsection{Proof of Proposition \ref{stat}}
We recall the
following facts on the permutation group.
\begin{itemize}
\item The decomposition into cycles leads to a partition of $S_n$ into equivalence classes
of permutations with the same cycle structure
$C_{\mathbf{q}}$, where $\mathbf{q} = [q_1, q_2, \dots ,q_r]$ is an
unordered $r$-tuple of natural numbers with $q_1 + q_2 + \dots + q_r =n$.
\item Two permutation $\pi'$ and $\pi''$ belong to the same class if and only if they are conjugate in
$S_n$, i.e. if there exists a $\pi \in S_n$ such that
\begin{equation} \label{conjclass}
    \pi'' = \pi^{-1} \pi' \pi.
\end{equation}
\item The number of permutations belonging to the class $C_{\mathbf{q}}$ is
\begin{equation} \label{countingperms}
    \frac{n!}{n_{\mathbf{q}}! (q_1 q_2\dots q_r)}
\end{equation}
with $n_\mathbf{q}! = n_1! n_2! \dots n_j! \dots$ and $n_j$ is the number of cycles of length $j$
in $\mathbf{q}$.
\end{itemize}

We observe that since our Hamiltonian is symmetric ($[\hcHam,U_\pi] = 0, \pi \in S_n$) and therefore for
$\pi', \pi'' \in C_{\mathbf{q}}$, one has by (\ref{conjclass})
\begin{equation}
  \begin{split}                                 \label{invar_under_cycles} 
    \trace_{\Hhccan{n}} \left[ U_{\pi''} \e^{-\beta \hcHam} \right]
        &= \trace_{\Hhccan{n}} \left[U_\pi^{-1} U_{\pi'} U_\pi \e^{-\beta \hcHam} \right] \\
        &= \trace_{\Hhccan{n}} \left[ U_\pi^{-1} U_{\pi'} \e^{-\beta \hcHam} U_\pi \right] \\
        &= \trace_{\Hhccan{n}} \left[ U_{\pi'} \e^{-\beta \hcHam} \right].
   \end{split}
\end{equation}


For $q \in \mathbb{N}$, let $N_q(\pi)$ be the number of cycles of length $q$ in $\pi$.

Let $r_j$ denote the number of cycles of length $j$. Then $\sum_{j} j r_j = n$ and the corresponding
number of permutations this cycle structure is $n!/ \prod_{j} j^{r_j} r_j!$ (from (\ref{countingperms})).
Denote $(r_j)$ the class of permutations with such a cycle structure.
Then
\begin{align*}
	\cProb(N_q\!=\!r)
&=	\frac{1}{\cPart} \frac{1}{n!}
	\mathop{\sum_{(r_j)} }_{ r_q = r} \sum_{\pi \in (r_j)}
	\trace_{\Hhccan{n}} \left[ U_\pi \e^{-\beta \hcHam} \right]
\\
&=	\frac{1}{\cPart} \frac{1}{n!}
	\mathop{\sum_{(r_j)}}_{ r_q = r}
	\frac{n!}{\prod_{j \ge 1} j^{r_j} r_j !}
	\trace_{\Hhccan{n}} \left[ U_{\tilde{\pi}} \e^{-\beta \hcHam} \right]
\end{align*}
where $\tilde{\pi}$ is any permutation with cycle distribution $(r_j)$. Suppose that $r \ge 1$
and consider a permutation where the first $q$ indices belong to the same cycle of length $q$.
Let $\pi'$ denote the permutation of the remaining $n-q$ indices. We have
\[
	U_\pi = U_q \otimes U_{\pi'}
\]
and $\pi'$ has cycle structure $(r_j - \delta_{jq})$. Then
\begin{align*}
	\cProb(N_q\!=\!r)
&=	\frac{1}{\cPart} \frac{1}{n!}
	\mathop{\sum_{(r_j), \sum j r_j = n-q}}_{ r_q = r-1}
	\frac{(n-q)!}{\prod_{j \ge 1} j^{r_j} r_j!} \frac{n!}{qr(n-q)!}
	\trace_{\Hhccan{n}} \left[ (U_q \otimes U_{\pi'}) \e^{-\beta \hcHam} \right]
\\
&=	\frac{1}{\cPart} \frac{1}{qr(n-q)!}
	\sum_{\pi' \in S_{n-q}}
	\trace_{\Hhccan{n}} \left[ (U_q \otimes U_{\pi'}) \e^{-\beta \hcHam} \right]
\end{align*}	

Then the canonical expectation of the number of $q$-cycles is found to be
\begin{align*}
    \cEx(N_q) &= \sum_{r=0}^{\infty} r\cProb(N_q = r)
\\
    &=	\frac{1}{\cPart} \frac{1}{q(n-q)!}
    		\sum_{\pi' \in S_{n-q}}
		\trace_{\Hhccan{n}} \left[ (U_q \otimes U_{\pi'}) \e^{-\beta \hcHam} \right] \hspace{1.4cm}
\end{align*}
\begin{align*}
\phantom{\cEx(N_q)}
    &=	\frac{1}{\cPart} \frac{1}{q(n-q)!} \sum_{\pi' \in S_{n-q}}
		\trace_{\Hcan{n}} \left[ \Phc (U_q \otimes U_{\pi'}) \e^{-\beta \hcHam} \Phc \right]
\\
    &=	\frac{1}{\cPart} \frac{1}{q}
		\trace_{\Hcan{q} \otimes \Hcansym{n-q}}
		\left[ \Phc (U_q \otimes I^{(n-q)}) \e^{-\beta \hcHam} \Phc \right]
\\
    &=	\frac{1}{\cPart} \frac{1}{q}
		\trace_{\Phc(\Hcan{q} \otimes \Hcansym{n-q})}
		\left[ (U_q \otimes I^{(n-q)}) \e^{-\beta \hcHam} \right].
\end{align*}

Since
\[
 	c_\sV^n(q) = \frac{q \; \cEx(N_q) }{V}
\]
we have proved Proposition \ref{stat}.
%
\subsection{Proof of Proposition \ref{c}}\label{sub c}
To prove Proposition \ref{c} we have to obtain an upper bound  for
\begin{equation*}
    \left |\trace_{\Hhccyclespace}
        \left[ U_q \e^{-\beta H^{\mathsf{hc}}_{n,\sV} } \right]  \\
    - \trace_{\Hhccyclespace}
        \left[ U_q \e^{-\beta \widetilde{H}^{\mathsf{hc}}_{q,n,\sV} } \right]\right |
\end{equation*}
In order to do this we first shall introduce some notation and make some remarks before
proceeding.
\par
Let $\Lambda_{\sV +}^{(n-q)}$ be the family of sets of $n-q$ distinct points
of $\Lambda_\sV$. For $\mathbf{k}=\{k_1, k_2, \dots , k_{n-q}\}\in \Lambda_{\sV +}^{(n-q)}$
let
\begin{equation*}\label{}
  |\mathbf{k}\ra := \sigma_{+}^{n-q}(\mathbf{e}_{k_1} \otimes \mathbf{e}_{k_2}
  \otimes \dots \otimes \mathbf{e}_{k_{n-q}}).
  \end{equation*}
Then $\{|\mathbf{k}\ra\,|\,\mathbf{k}\in \Lambda_{\sV +}^{(n-q)}\}$
is an orthonormal basis for
$\mathcal{H}_{(n-q),\sV, +}^{\mathrm{hc}}
:=\mathcal{P}^{\mathrm{hc}}_{n-q} \mathcal{H}_{\sV +}^{(n-q)}$.
\par
Similarly let $\Lambda_{\sV}^{(q)}$ be the set of
ordered $q$-tuples of distinct indices of $\Lambda_\sV$ and
for $\mathbf{i}=(i_1, i_2, \dots , i_q)\in \Lambda_{\sV}^{(q)}$
let
\begin{equation*}\label{}
  |\mathbf{i}\ra := \mathbf{e}_{i_1} \otimes \mathbf{e}_{i_2}
  \otimes \dots \otimes \mathbf{e}_{i_q}.
  \end{equation*}
Then $\{|\mathbf{i}\ra\,|\,\mathbf{i}\in \Lambda_{\sV}^{(q)}\}$ is an orthonormal basis for
$\mathcal{H}_{q,\sV}^{\mathrm{hc}}:=\mathcal{P}^{\mathrm{hc}}_q \mathcal{H}_{\sV}^{(q)}$.
\par
If $\mathbf{k}\in \Lambda_{\sV +}^{(n-q)}$ and
$\mathbf{i}\in \Lambda_{\sV}^{(q)}$
we shall write $\mathbf{k}\sim\mathbf{i}$
if $\{k_1, k_2, \dots , k_{n-q}\}\cap \{i_1, i_2, \dots , i_q\}=\emptyset$
and we shall use the notation
\begin{equation*}\label{}
    |\mathbf{i};\mathbf{k}\ra:=|\mathbf{i}\ra\otimes|\mathbf{k}\ra.
\end{equation*}
Then a basis for $\Hhccyclespace$ may
be formed by taking the tensor product of the bases
of $\mathcal{H}_{(n-q),\sV, +}^{\mathrm{hc}}$ and
$\mathcal{H}_{q,\sV}^{\mathrm{hc}}$ where we disallow particles from appearing in both
spaces simultaneously.
Thus the set
$\{|\mathbf{i};\mathbf{k}\ra\,|\,\mathbf{k}\in \Lambda_{\sV +}^{(n-q)},\
\mathbf{i}\in \Lambda_{\sV}^{(q)},\ \mathbf{k}\sim\mathbf{i}\}$
is an orthonormal basis for $\Hhccyclespace$.
\par
We shall need also the following facts. For simplicity we shall write $\Htilde$ and
$\widetilde{P}$ for $\widetilde{H}^{\mathsf{hc}}_{q,n,\sV}$ and $\widetilde{P}^{(q)}_\sV$
respectively, as defined in equation (\ref{redHam}).
\par
Let $\mathcal{P}_\ii^{(n-q)}$ be the projection
of $\mathcal{H}_{(n-q),\sV, +}^{\mathrm{hc}}$ onto a space
with none of the $n-q$ particles at the points $i_1, i_2, \dots , i_q$
(so there are $V-q$ available sites for $n-q$ particles) and not more than
one particle at any site. Then

\begin{remark}\label{rem a}
For $\ii \sim \kk$, if $s>0$
\begin{equation}\label{Htilde}
	\e^{-\beta \Htilde s} | \ii; \kk \ra = |\ii; \e^{-\beta H^\ii s} \kk \ra \e^{-\beta q s}
\end{equation}
where $H^\ii = \mathcal{P}_\ii^{(n-q)} ((n-q) - P_\sV^{(n-q)} ) \mathcal{P}_\ii^{(n-q)}$.
\par
\end{remark}
This can be seen as follows:
For $\ii \sim \kk$,
\begin{align*}
    \widetilde{H} |\ii; \kk \ra
&=  \Phc ( n - P_\sV^{(n-q)} ) \Phc |\ii; \kk \ra
\\
&=  q\Phc |\ii; \kk \ra + \Phc | \ii; ( (n-q)- P_\sV^{(n-q)} ) \kk \ra
\\
&=  q |\ii; \kk \ra +  |\ii; \mathcal{P}_\ii^{(n-q)} ( (n-q) - P_\sV^{(n-q)} ) \kk \ra
\\
&=  q |\ii; \kk\ra +  |\ii; H^\ii\kk \ra.
\end{align*}
\begin{remark}\label{rem b}
For $\ii \sim \kk$,
\begin{equation}\label{}
	H^\ii|\kk \ra=(n-q)|\kk \ra-\frac{1}{V}
		 \sum_{l=1}^{n-q}\sum_{j\notin\, \ii\,\cup\,\kk\setminus\{k_l\}}
		 |(k_1,k_2, \ldots,\widehat{k}_l,j,\ldots,k_{n-q}) \ra
\end{equation}
where the hat symbol implies the term is removed from the sequence,
while from (\ref{h-lambda}), for $\kk\in \mathcal{H}^{(n-q)}_{\sV-q,+}$ we have
\begin{equation}
	H^{\mathsf{hc}}_{\lambda, n-q,\sV-q} | \kk \ra = (n-q)|\kk \ra-\frac{\lambda}{V-q}
    		\sum_{l=1}^{n-q}
		\sum^{V-q}_{\stackrel{j=1}{j\notin\, \kk\setminus\{k_l\}}}
		|(k_1,k_2, \ldots,\widehat{k}_l,j,\ldots,k_{n-q}) \ra.
\end{equation}
Thus $H^\ii$ is unitarily equivalent to $H^{\mathsf{hc}}_{(\sV-q)/\sV,\, n-q,\sV-q}$ and
\begin{equation}
	\trace_{\Hhccyclespace}  \left[
		\mathcal{P}_\ii^{(n-q)} \e^{-\beta \widetilde{H}^\ii}  \mathcal{P}_\ii^{(n-q)}
	\right]
    	=\ccPart.
\end{equation}
\end{remark}
\begin{remark}\label{rem c}
For $s, \alpha \in \RR$
    \[
        \left( \mathcal{P}_\ii^{(n-q)} \e^{-s \widetilde{H}^\ii}  \mathcal{P}_\ii^{(n-q)} \right)^\alpha
        = \mathcal{P}_\ii^{(n-q)} \e^{-s \alpha \widetilde{H}^\ii}  \mathcal{P}_\ii^{(n-q)}.
    \]
\end{remark}
We expand
\begin{equation*}
    \trace_{\Hhccyclespace} \left[U_q \e^{-\beta H^{\mathsf{hc}}_{n,\sV} } \right]
    = \trace_{\Hhccyclespace} \left[U_q \e^{-\beta (\Htilde - \widetilde{P}) } \right]
\end{equation*}
in a Dyson series in powers of
$\widetilde{P}$. If $m\geq 1$, the $m^\text{th}$ term is
\begin{align*}
    X_m := \beta^m \int_0^1\hskip -0.3cm ds_1 \int_0^{s_1} \hskip -0.4cmds_2 \dots
    & \int_0^{s_{m-1}}\hskip -0.8cm ds_m  \;\;
    \trace_{\mathcal{H}^{\mathsf{hc}}_{q,n,\sV}}
        \Bigg[ \e^{-\beta \Htilde (1-s_1)} \widetilde{P}
        \e^{-\beta \Htilde (s_1-s_2)} \widetilde{P}
        \cdots \\
        & \hskip 6cm\cdots \widetilde{P}
        \e^{-\beta \Htilde (s_{m-1} -s_m)} \widetilde{P}
        \e^{-\beta \Htilde s_m}  U_q \Bigg].
\end{align*}
Recall that $\widetilde{P}:= \Phc P_\sV^{(q)} \Phc$ where
\begin{equation}\label{}
P_\sV^{(q)} =
P_\sV \otimes I \otimes \cdots \otimes I + I \otimes P_\sV \otimes I \otimes \cdots \otimes I + \cdots +
I \otimes \cdots \otimes I \otimes P_\sV
\end{equation}
has $q$ terms, so in the above trace we have $m$ instances of this form.
Let $P^{(q)}_r = I \otimes \cdots \otimes
\underbrace{P_\sV}_{\text{$r^{th}$ place}} \otimes \dots \otimes I$, and let
$\widetilde{P}_r = \Phc P^{(q)}_r \Phc$.
\par
Then we can write
\begin{align*}
    X_m &= \beta^m \int_0^1 \hskip -0.3cm ds_1 \int_0^{s_1}\hskip -0.4cm ds_2 \dots
    \int_0^{s_{m-1}}\hskip -0.8cm ds_m \;\; \sum_{r_1 = 1}^q \cdots \sum_{r_m = 1}^q \\
    & \qquad \times \trace_{\mathcal{H}^{\mathsf{hc}}_{q,n,\sV}}
        \Bigg[ \e^{-\beta \Htilde (1-s_1)} \widetilde{P}_{r_1}
        \e^{-\beta \Htilde (s_1-s_2)} \widetilde{P}_{r_2}
        \cdots \widetilde{P}_{r_{m-1}}
        \e^{-\beta \Htilde (s_{m-1} -s_m)} \widetilde{P}_{r_m}
        \e^{-\beta \Htilde s_m}  U_q \Bigg].
\end{align*}
In terms of the basis of $\mathcal{H}^{\mathsf{hc}}_{q,n,\sV}$ we may write the expression for $X_m$ as
\begin{multline}                            \label{trace_term}   
    \hspace{1cm} X_m=\beta^m \int_0^1\hskip -0.3cm ds_1 \int_0^{s_1} \hskip -0.4cm ds_2
    \dots  \int_0^{s_{m-1}}\hskip -0.8cm ds_m \;\;
    \sum_{r_1=1}^q \dots \sum_{r_m=1}^q \;\;
    \sum_{\kk^0, \dots\ ,\kk^m} \;\; \sum_{\ii^0 \sim \kk^0} \cdots
    \sum_{\ii^m \sim \kk^m} \\
        \la \ii^0; \kk^0 | \e^{-\beta \Htilde (1-s_1)} \widetilde{P}_{r_1}
        | \ii^1; \kk^1 \ra
        \la \ii^1; \kk^1 | \e^{-\beta \Htilde (s_1-s_2)} \widetilde{P}_{r_2}
        | \ii^2; \kk^2 \ra \cdots\\
        \cdots
        \la \ii^{m-1}; \kk^{m-1} |\e^{-\beta \Htilde (s_{m-1} -s_m)} \tilde{P}_{r_m}
        | \ii^m; \kk^m \ra
        \la \ii^m; \kk^m |\e^{-\beta \Htilde s_m}  U_q  | \ii^0; \kk^0 \ra
\end{multline}
where it is understood that the $\ii$ summations are over $\Lambda_{\sV}^{(q)}$
and the $\kk$ summations are over $\Lambda_{\sV +}^{(n-q)}$.
Note that for $\ii \sim \kk$
\begin{equation}\label{P-r}
	\widetilde{P}_r | \ii ; \kk \ra = \frac{1}{V}\hskip -0.8cm \sum_{\stackrel{l=1...V}
		{l \notin \kk;\ l \neq i_1 \dots \widehat{i}_r \dots i_q}}\hskip -0.8cm
		|(i_1,\cdots , \widehat{i}_r ,l , \cdots , i_q ); \kk \ra
\end{equation}
where again the hat symbol implies that the term is removed from the sequence.
\par
Consider one of the inner products in the expression (\ref{trace_term}) for $X_m$, using
(\ref{Htilde}) and (\ref{P-r}) above. For
$\ii \sim \kk$ and $\jj \sim \kk'$:
\begin{align*}
	\la \ii ; \kk\, |\, \e^{-\beta s \widetilde{H}} \widetilde{P}_r | \jj ; \kk' \ra
&= 	\frac{\e^{-\beta q s}}{V}\hskip -0.7cm  \sum_{\stackrel{l=1...V}{l \notin \kk';\, l \neq j_1, \dots
	\widehat{j_r}, \dots j_q}} \hskip -0.7cm\la \ii\,|\,( j_1,
	\cdots ,
	\widehat{j_r}, l, \cdots ,j_q) \ra
	\la \kk\, |\, \e^{-\beta s H^\ii} | \kk' \ra
\\[0.3cm]
&= 	\frac{\e^{-\beta q s}}{V}
	\hskip -0.7cm\sum_{\stackrel{l=1...V}{l \notin \kk';\, l \neq j_1, \dots ,
	\widehat{j_r} , \dots ,  j_q}}\hskip -0.7cm \delta_{i_1 j_1}
	\dots \widehat{\delta_{i_r j_r}} \, \delta_{i_r l} \,
	\dots \delta_{i_q j_q}
	\la \kk | \e^{-\beta s H^\ii} | \kk' \ra.
\end{align*}
In summing over $l$ we replace $l$ by $i_r$ and the result is non-zero
only if $i_r \notin \kk'$ and $i_r \neq j_1, \dots , \widehat{j_r},  \dots , j_q$.
However this last condition is not necessary because if $i_r=j_s\ (s\neq r)$
then $j_s\neq i_s$ and we get zero. Also if for some $s\neq r$, $i_s \in \kk'$ then once again
$j_s\neq i_s$. We can therefore replace the condition $i_r \notin \kk'$ by $\ii \sim \kk'$.
Using $\mathcal{I}$ for the indicator function, we have
\begin{align*}
	\la \ii ; \kk\, |\, \e^{-\beta s \widetilde{H}} \widetilde{P}_r | \jj ; \kk' \ra
&= \frac{\e^{-\beta q s}}{V}  \delta_{i_1 j_1}
    	\dots \widehat{\delta_{i_r j_r}} \, \dots \delta_{i_q j_q}
	\la \kk | \e^{-\beta s H^\ii} | \kk' \ra
	\mathcal{I}_{(\ii \sim \kk')}
\\[0.2cm]
&= 	\frac{\e^{-\beta q s}}{V}  \delta_{i_1 j_1} \dots
	\widehat{\delta_{i_r j_r}} \, \dots \delta_{i_q j_q}
	\la \kk | \mathcal{P}_\ii^{(n-q)} \e^{-\beta s H^\ii}
	\mathcal{P}_\ii^{(n-q)} | \kk' \ra
\end{align*}
Now if we sum over $\jj \sim \kk'$, with $\ii \sim \kk$ and for a fixed $r$:
\begin{align*}
	\sum_{ \jj \sim \kk'} \la \ii ; \kk | \e^{-\beta \widetilde{H} s}
	\widetilde{P}_r | \jj ; \kk' \ra \la \jj ; \kk' |
&= 	\frac{\e^{-\beta s q}}{V}  \la \kk | \mathcal{P}_\ii^{(n-q)}
	\e^{-\beta H^\ii s} \mathcal{P}_\ii^{(n-q)} | \kk' \ra \\
&	\quad \times \hspace{-0.4cm}
	\sum_{\stackrel{j_r = 1 \dots V}{j_r \notin \kk' \cup \ii\setminus \{ i_r \}}}
	\la (i_1, \dots ,i_{r-1}, j_r, i_{r+1}, \dots,i_q); \kk' |  .
\end{align*}
It is convenient to define the operation $[r,x](\ii)$ which inserts the value of $x$
in the $r^\text{th}$ position of $\ii$ instead of $i_r$.
So for example taking the ordered triplet $\ii = (5,4,1)$,
then $[2,8](\ii) = (5,8,1)$. For brevity we shall denote the composition of these
operators as $[r_k, x_k; \, \dots \, ; r_2, x_2 ; r_1, x_1] := [r_k, x_k]  \circ \cdots \circ [r_2, x_2] \circ [r_1, x_1]$.

Thus the final term in the above expression may be rewritten as $\la [r, j_r](\ii) ; \kk' | $.
\par
Performing two summations for fixed $r_1$ and $r_2$ we get:
\begin{align*}
    \sum_{ \ii^1 \sim \kk^1} \sum_{ \ii^2 \sim \kk^2} &
    \la \ii^0 ; \kk^0 |
    \e^{-\beta s\widetilde{H} } \widetilde{P}_{r_1} | \ii^1 ; \kk^1 \ra
    \la \ii^1 ; \kk^1 | \e^{-\beta t\widetilde{H} } \widetilde{P}_{r_2}
    | \ii^2 ; \kk^2 \ra
    \la \ii^2 ; \kk^2 |  \\
&   =  \frac{\e^{-\beta q (s+t)}}{V^2}
    \sum_{i^1_{r_1} \notin \kk^1 \cup \ii^0 \setminus \{ i^0_{r_1} \}} \hspace{0.4cm}
    \sum_{i^2_{r_2} \notin \kk^2 \cup [r_1,i^1_{r_1}](\ii^0) \setminus \{ i^0_{r_2} \}}
    \la \kk^0 | \mathcal{P}_{\ii^0}\
    \e^{-\beta sH^{\ii^0} }\ \mathcal{P}_{\ii^0} | \kk^1 \ra \\
&  \qquad \times\la \kk^1 | \mathcal{P}_{[r_1,i^1_{r_1}](\ii^0)}\
\e^{-\beta t H^{[r_1,i^1_{r_1}](\ii^0)} }
   \  \mathcal{P}_{[r_1,i^1_{r_1}](\ii^0)} | \kk^2 \ra
    \la [r_1, i^1_{r_1}; r_2, i^2_{r_2}](\ii^0); \kk^2 |  \\
& \\
&   =  \frac{\e^{-\beta q (s+t)}}{V^2}
    \sum_{i^1_{r_1} \notin  \ii^0 \setminus \{ i^0_{r_1} \}} \hspace{0.4cm}
    \sum_{i^2_{r_2} \notin \kk^2 \cup [r_1,i^1_{r_1}](\ii^0) \setminus \{ i^0_{r_2} \}}
    \la \kk^0 |\ \mathcal{P}_{\ii^0}\ \e^{-\beta s H^{\ii^0} }\ \mathcal{P}_{\ii^0}\
    | \kk^1 \ra
\\
&   \qquad \times \la \kk^1 |\ \mathcal{P}_{[r_1,i^1_{r_1}](\ii^0)}\ \e^{-\beta t H^{[r_1,i^1_{r_1}](\ii^0)}}
   \ \mathcal{P}_{[r_1,i^1_{r_1}](\ii^0)}\ | \kk^2 \ra
    \la [r_2, i^2_{r_2} ; r_1, i^1_{r_1}] (\ii^0); \kk^2|
\\
\end{align*}
due to the fact that $\mathcal{P}_{[r_1, i^1_{r_1}](\ii^0)}|\kk^1\ra=0$ if $i^1_{r_1}\in \kk^1$.
We may apply this to all inner product terms of (\ref{trace_term}) except the final one.
Note we sum over the $V$ sites of the lattice, with certain points excluded in each case.

For the final inner product of (\ref{trace_term}) we obtain:
\begin{align*}
& \la [r_m, i^m_{r_m}; \,  \dots \, ; r_2, i^2_{r_2} ; r_1, i^1_{r_1}](\ii^0) ; \kk^m |
        \e^{-\beta s_m \widetilde{H}  } U_q | \ii^0 ; \kk^0 \ra
\\[0.2cm]
&=  \e^{-\beta q s_m} \la [r_m, i^m_{r_m}; \, \dots \, ; r_2, i^2_{r_2} ; r_1, i^1_{r_1}](\ii^0); \kk^m |
        \e^{-\beta  s_m H^{[r_m, i^m_{r_m}; \, \dots \, ; r_2, i^2_{r_2} ; r_1, i^1_{r_1}](\ii^0)} }
        U_q | \ii^0 ; \kk^0 \ra
\\[0.2cm]
&=  \e^{-\beta q s_m} \la \kk^m |
        \e^{-\beta  s_m H^{[r_m, i^m_{r_m}; \, \dots \, ; r_2, i^2_{r_2} ; r_1, i^1_{r_1}](\ii^0)} }
        | \kk^0 \ra \;
        \la [r_m, i^m_{r_m}; \, \dots \, ; r_2, i^2_{r_2} ; r_1, i^1_{r_1}](\ii^0) | U_q \ii^0 \ra
\\[0.2cm]
&=  \e^{-\beta q s_m} \la \kk^m |
        \mathcal{P}_{[r_m, i^m_{r_m}; \, \dots \, ; r_2, i^2_{r_2} ; r_1, i^1_{r_1}](\ii^0)}
        \e^{-\beta s_m H^{[r_m, i^m_{r_m}; \, \dots \, ; r_2, i^2_{r_2} ; r_1, i^1_{r_1}](\ii^0)}  }
        \mathcal{P}_{[r_m, i^m_{r_m}; \, \dots \, ; r_2, i^2_{r_2} ; r_1, i^1_{r_1}](\ii^0) }
        | \kk^0 \ra
\\
    & \hspace{4cm} \times  \la [r_m, i^m_{r_m};  \, \dots \, ; r_2, i^2_{r_2} ; r_1, i^1_{r_1}](\ii^0) | U_q \ii^0 \ra . \\
\end{align*}
Applying this to the whole tracial expression of (\ref{trace_term}) we obtain
\begin{align*}
   	X_m
=& \; \e^{-\beta q} \frac{\beta^m}{V^m} \sum_{\kk^0 \dots \kk^m} \sum_{\ii^0}
	\sum_{i^1_{r_1} \notin \ii^0 \setminus \{i^0_{r_1}\}} \hspace{0.3cm}
	\sum_{i^2_{r_2} \notin [r_1,i^1_{r_1}](\ii^0) \setminus \{i^1_{r_2}\}}
	\cdots \sum_{i^m_{r_m} \notin [r_{m-1}, i^{m-1}_{r_{m-1}}; \, \dots \, ; r_2, i^2_{r_2} ; r_1, i^1_{r_1}](\ii^0)
	\setminus \{i^{m-1}_{r_m}\}}
\\
&	\la \kk^0 | \mathcal{P}_{\ii^0} \e^{-\beta(1-s_1) \Htilde^{\ii^0} } \mathcal{P}_{\ii^0}
	| \kk^1 \ra
\\
&	\la \kk^1 | \mathcal{P}_{[r_1,i^1_{r_1}](\ii^0)}
	\e^{-\beta (s_1-s_2)\Htilde^{[r_1,i^1_{r_1}](\ii^0)}}
	\mathcal{P}_{[r_1,i^1_{r_1}](\ii^0)} | \kk^2 \ra
\\
&	\la \kk^2 | \mathcal{P}_{[r_2, i^2_{r_2}; r_1, i^1_{r_1}](\ii^0)} \e^{-\beta (s_2-s_3)
	\Htilde^{[r_2, i^2_{r_2} ; r_1, i^1_{r_1}](\ii^0)} }
	\mathcal{P}_{[r_2, i^2_{r_2} ; r_1, i^1_{r_1}](\ii^0)} | \kk^3 \ra
\\
&	\cdots
\\
&	\la \kk^m | \mathcal{P}_{[r_m, i^m_{r_m}; \, \dots \, ; r_2, i^2_{r_2} ; r_1, i^1_{r_1}](\ii^0)}
	\e^{-\beta s_m \Htilde^{ [r_m, i^m_{r_m}; \, \dots \, ; r_2, i^2_{r_2} ; r_1, i^1_{r_1}](\ii^0)} }
	\mathcal{P}_{[r_m, i^m_{r_m}; \, \dots \, ; r_2, i^2_{r_2} ; r_1, i^1_{r_1}](\ii^0)}
	| \kk^0 \ra \\
&   \la [r_m, i^m_{r_m}; \, \dots \, ; r_2, i^2_{r_2} ; r_1, i^1_{r_1}](\ii^0) | U_q \ii^0 \ra
\\[0.4cm]
=& 	\;\e^{-\beta q} \frac{\beta^m}{V^m} \sum_{\ii^0}
	\sum_{i^1_{r_1} \notin \ii^0 \setminus \{i^0_{r_1}\}} \hspace{0.3cm}
	\sum_{i^2_{r_2} \notin [r_1,i^1_{r_1}](\ii^0) \setminus \{i^1_{r_2}\}}
	\cdots \sum_{i^m_{r_m} \notin [r_{m-1}, i^{m-1}_{r_{m-1}}; \, \dots \, ; r_2, i^2_{r_2} ; r_1, i^1_{r_1}](\ii^0)
	\setminus \{i^{m-1}_{r_m}\}}
\\
&	\la [r_m, i^m_{r_m}; \, \dots \, ; r_2, i^2_{r_2} ; r_1, i^1_{r_1}](\ii^0) | U_q \ii^0 \ra
\\
&	\trace_{\mathcal{H}_{(n-q),\sV, +}^{\mathrm{hc}}} \Bigg[ \mathcal{P}_{\ii^0} \e^{-\beta (1-s_1)\Htilde^{\ii^0} }
	\mathcal{P}_{\ii^0} \mathcal{P}_{i^1_{r_1}(\ii^0)}
	\e^{-\beta (s_1-s_2)\Htilde^{i^1_{r_1}(\ii^0)} }
	\mathcal{P}_{i^1_{r_1}(\ii^0)} \cdots
\\
&	\qquad \cdots \mathcal{P}_{[r_m, i^m_{r_m}; \, \dots \, ; r_2, i^2_{r_2} ; r_1, i^1_{r_1}](\ii^0)}
	\e^{-\beta s_m\Htilde^{[r_m, i^m_{r_m}; \, \dots \, ; r_2, i^2_{r_2} ; r_1, i^1_{r_1}](\ii^0)} }
	\mathcal{P}_{[r_m, i^m_{r_m}; \, \dots \, ; r_2, i^2_{r_2} ; r_1, i^1_{r_1}](\ii^0)}  \Bigg].
\end{align*}
From the H\"older inequality (see Manjegani \cite{Manjegani}), for finite dimensional non-negative matrices
$A_1, A_2, \dots , A_{m+1}$ we have the inequality
\[
    \left| \trace \big( A_1 A_2 \dots A_{m+1} \big) \right|
        \le \trace \big| A_1 A_2 \dots A_{m+1} \big|
        \le \prod_{k=1}^{m+1} \big( \trace A_k^{p_k} \big)^{\tfrac{1}{p_k}}
\]
where $\sum_{k=1}^{m+1} \tfrac{1}{p_k} = 1$, $p_i > 0$.

Set $p_1 = \frac{1}{1-s_1},\ p_2 = \frac{1}{s_1 - s_2},\ \dots ,\ p_m = \frac{1}{s_{m-1}-s_m},\ p_{m+1} = \frac{1}{s_m}$.
Taking the modulus of the above trace
\begin{align*}
&	\Bigg| \trace_{\mathcal{H}_{(n-q),\sV, +}^{\mathrm{hc}}}  \Bigg[ \mathcal{P}_{\ii^0}
	\e^{-\beta \Htilde^{\ii^0} (1-s_1)} \mathcal{P}_{\ii^0}
	\mathcal{P}_{[r_1, i^1_{r_1}](\ii^0)} \e^{-\beta \Htilde^{[r_1, i^1_{r_1}](\ii^0)} (s_1-s_2)}
	\mathcal{P}_{[r_1, i^1_{r_1}](\ii^0)} \; \cdots \\
&	\qquad \qquad \qquad \cdots \mathcal{P}_{[r_m, i^m_{r_m}; \, \dots \, ; r_2, i^2_{r_2} ; r_1, i^1_{r_1}](\ii^0)}
	\e^{-\beta \Htilde^{[r_m, i^m_{r_m}; \, \dots \, ; r_2, i^2_{r_2} ; r_1, i^1_{r_1}](\ii^0)} (s_m)}
	\mathcal{P}_{[r_m, i^m_{r_m}; \, \dots \, ; r_2, i^2_{r_2} ; r_1, i^1_{r_1}](\ii^0)}
	\Bigg]    \Bigg|
\\
&	\le \quad \trace_{\mathcal{H}_{(n-q),\sV, +}^{\mathrm{hc}}} \bigg[ \mathcal{P}_{\ii^0}
	\e^{-\beta \Htilde^{\ii^0}} \mathcal{P}_{\ii^0} \bigg]^{1-s_1}
	\trace_{\Hhccan{n-q}} \bigg[ \mathcal{P}_{[r_1, i^1_{r_1}](\ii^0)}
	\e^{-\beta \Htilde^{[r_1, i^1_{r_1}](\ii^0)}} \mathcal{P}_{[r_1, i^1_{r_1}](\ii^0)} \bigg]^{s_1-s_2} \cdots \\
&	\quad \cdot\!\cdot\!\cdot\! \trace_{\mathcal{H}_{(n-q),\sV, +}^{\mathrm{hc}}}
	\bigg[ \mathcal{P}_{[r_m, i^m_{r_m}; \, \dots \, ; r_2, i^2_{r_2} ; r_1, i^1_{r_1}](\ii^0)}
	\e^{-\beta \Htilde^{[r_m, i^m_{r_m}; \, \dots \, ; r_2, i^2_{r_2} ; r_1, i^1_{r_1}](\ii^0)}}
	\mathcal{P}_{ [r_m, i^m_{r_m}; \, \dots \, ; r_2, i^2_{r_2} ; r_1, i^1_{r_1}](\ii^0)} \bigg]^{s_m} \!\!\!.
\end{align*}
Since the trace is independent of the $V-q$ sites
$\{\ii^0, [r_1,i^1_{r_1}](\ii^0),\,  \dots, \, [r_m, i^m_{r_m}; \, \dots\\ \, ; r_2, i^2_{r_2} ; r_1, i^1_{r_1}](\ii^0)\}$,
and therefore using Remark \ref{rem c}, the product of all the trace terms above is equal to
\[
	\trace_{\mathcal{H}_{(n-q),\sV, +}^{\mathrm{hc}}} \left[ \mathcal{P}_{\boldl}
		\e^{-\beta \Htilde^{\boldl}} \mathcal{P}_{\boldl} \right]
\]
with $\boldl = \{V-q+1, V-q+2, \dots ,V\}$ and from Remark \ref{rem b},
\begin{equation}
	\trace_{\mathcal{H}_{(n-q),\sV, +}^{\mathrm{hc}}} \left[ \mathcal{P}_{\boldl}
		\e^{-\beta \Htilde^{\boldl}} \mathcal{P}_{\boldl} \right]
	=\ccPart.
\end{equation}
\par
Consider the sum
\begin{multline}  					\label{sum_of_cycled_inner_products} 
	\sum_{\ii^0} \; \sum_{i^1_{r_1} \notin \ii^0 \setminus \{i^0_{r_1}\}} \hspace{0.2cm}
	\sum_{i^2_{r_2} \notin [r_1, i^1_{r_1}](\ii^0) \setminus \{i^1_{r_2}\}}
	\cdots
\\
	\cdots \sum_{i^m_{r_m} \notin [r_{m-1}, i^{m-1}_{r_{m-1}}; \, \dots \, ; r_2, i^2_{r_2} ; r_1, i^1_{r_1}](\ii^0)
	\setminus \{i^{m-1}_{r_m}\}}  \hspace{-0.5cm}
	\la [r_m, i^m_{r_m}; \, \dots \, ; r_2, i^2_{r_2} ; r_1, i^1_{r_1}](\ii^0) | U_q \ii^0 \ra.
\end{multline}

If $\{r_1, r_2, \dots , r_m\} \ne \{1,2,\dots, q\} $,  then  $| [r_m, i^m_{r_m}; \, \dots \, ; r_2, i^2_{r_2} ; r_1, i^1_{r_1}](\ii^0) \ra$
is of the form
\[
	| j_1, j_2, \dots ,j_{n_1},i^0_{n_1+1}, \dots ,i^0_{n_2}, j_{n_2+1},
	\dots, j_{n_3}, i^0_{n_3+1}, \dots  ,
	i^0_{n_4}, j_{n_4+1}, \dots \dots\ra
\]
where $\{n_1, n_2, \dots \}$ is a non-empty ordered set of distinct integers between 0 and $q$. This state is
clearly orthogonal to $U_q \ii^0$ for any $q$. Note that this situation does not arise if $q=1$.
Note also that  this is always the case if $ m < q$.
\par
We may bound the remaining sum corresponding to terms for which
$\{r_1, r_2, \dots , r_m\} = \{1,2,\dots, q\}$ by
\[
	\le \sum_{\ii^0} \;\;
	\underbrace{ \sum_{i^1_{r_1}=1}^\sV \;\;\;
	\sum_{i^2_{r_2}=1}^\sV \;\;\;
	\cdots  \;\;\;
	\sum_{i^m_{r_m} = 1}^\sV }_{\stackrel{\text{where $[r_m, i^m_{r_m}; \, \dots \,
		; r_2, i^2_{r_2} ; r_1, i^1_{r_1}]$}}{\text{\tiny{has distinct indices}}}} \;\;
	\la [r_m, i^m_{r_m}; \, \dots \, ; r_2, i^2_{r_2} ; r_1, i^1_{r_1}](\ii^0) | U_q \ii^0 \ra .
\]
Observe that in this case $| [r_m, i^m_{r_m}; \, \dots \, ; r_2, i^2_{r_2} ; r_1, i^1_{r_1}](\ii^0) \ra$ is independent of
$\ii^0$ so we may take it to be
\[
	|[r_m, i^m_{r_m}; \, \dots \, ; r_2, i^2_{r_2} ; r_1, i^1_{r_1}](\mathbf{s}^0) \ra
\]
where $\mathbf{s}^0 = (1,2,3,\dots,q)$. Then we can interchange the $\ii^0$ summation with the others, and
for each choice of $i^1_{r_1}, i^2_{r_2}, \dots , i^m_{r_m}$ there exists only one possible $\ii^0 \in \Lambda_\sV^{(q)}$
such that
\[
	\la [r_m, i^m_{r_m}; \, \dots \, ; r_2, i^2_{r_2} ; r_1, i^1_{r_1}](\ii^0) | U_q \ii^0 \ra \ne 0
\]
So we may conclude that
\begin{multline}
	\sum_{\ii^0} \; \sum_{i^1_{r_1} \notin \ii^0 \setminus \{i^0_{r_1}\}} \hspace{0.2cm}
	\sum_{i^2_{r_2} \notin [r_1, i^1_{r_1}](\ii^0) \setminus \{i^1_{r_2}\}}
	\cdots
\\
	\cdots \sum_{i^m_{r_m} \notin [r_{m-1}, i^{m-1}_{r_{m-1}}; \, \dots \, ; r_2, i^2_{r_2} ; r_1, i^1_{r_1}](\ii^0)
	\setminus \{i^{m-1}_{r_m}\}}  \hspace{-0.5cm}
	\la [r_m, i^m_{r_m}; \, \dots \, ; r_2, i^2_{r_2} ; r_1, i^1_{r_1}](\ii^0) | U_q \ii^0 \ra \le V^m.
\end{multline}

\;

Applying this, we see that the modulus of the integrated $m^\text{th}$ term of the
Dyson series may bounded above by
\begin{align*}
    |X_m| & \le \e^{-\beta q} \frac{\beta^m}{ m!} \frac{1}{V^m}\
    \ccPart\sum_{r_1 = 1}^q \cdots \sum_{r_m = 1}^q
\\
&   \qquad \times  \sum_{\ii^0} \;
	\sum_{i^1_{r_1} \notin \ii^0 \setminus \{i^0_{r_1}\}} \hspace{0.2cm}
	\sum_{i^2_{r_2} \notin [r_1, i^1_{r_1}](\ii^0) \setminus \{i^1_{r_2}\}}
	\cdots
\\
&	\qquad \cdots \sum_{i^m_{r_m} \notin [r_{m-1}, i^{m-1}_{r_{m-1}}; \, \dots
		 \, ; r_2, i^2_{r_2} ; r_1, i^1_{r_1}](\ii^0)
	\setminus \{i^{m-1}_{r_m}\}}  \hspace{-0.5cm}
	\la [r_m, i^m_{r_m}; \, \dots \, ; r_2, i^2_{r_2} ; r_1, i^1_{r_1}](\ii^0) | U_q \ii^0 \ra.
\\
&\le	\e^{-\beta q} \frac{\beta^m}{m!} \frac{1}{V^m}\
	\ccPart  \sum_{r_1 = 1}^q \cdots \sum_{r_m = 1}^q V^m
\\
&=	\e^{-\beta q} \frac{q^m \beta^m}{m!}\ccPart.
\end{align*}
Noting that the zeroth term of the Dyson series is
\[
	X_0 = \trace_{\mathcal{H}^{\mathsf{hc}}_{q,n,V}} \left[U_q \e^{-\beta \Htilde } \right],
\]
we may re-sum the series to obtain
\[
    \Bigg| \trace_{\mathcal{H}^{\mathsf{hc}}_{q,n,V}}
            \left[U_q \e^{-\beta H^{\mathsf{hc}}_{n,V} } \right]
        - \trace_{\mathcal{H}^{\mathsf{hc}}_{q,n,V}}
            \left[U_q \e^{-\beta \Htilde } \right]  \Bigg|
\le \e^{-\beta q}\ccPart\ \sum_{m=1}^\infty \frac{q^m \beta^m}{m!}.
\]
Thus
\begin{align*}
   \left|c_\sV^n(q)-\widetilde{c}_\sV^n(q)\right |
   & = \frac{1}{V} \left| \frac{
        \trace_{\mathcal{H}^{\mathsf{hc}}_{q,n,V}}
            \left[U_q \e^{-\beta H^{\mathsf{hc}}_{n,V} } \right]
        - \trace_{\mathcal{H}^{\mathsf{hc}}_{q,n,V}}
            \left[U_q \e^{-\beta \Htilde } \right]}{Z_\sbeta(n,V)} \right|
\\
    & \le \frac{\e^{-\beta q}}{V} \,
    \frac{ \ccPart}{Z_\sbeta(n,V)} \sum_{m=1}^\infty \frac{q^m \beta^m}{m!}
\\
    & = \frac{\e^{-\beta q}}{V} (\e^{\beta q} - 1)
    \frac{\ccPart}{Z_\sbeta(n,V)}.
    \end{align*}
\subsection{Proof of Proposition \ref{Z}}\label{sub Z}
Recall that we have
\begin{equation}\label{Z1}
	Z_\sbeta(n-q,V-q)
	= \trace_{\mathcal{H}^{\mathrm{hc}}_{n-q,\sV-q,+}} [\e^{-\beta H_{n-q,\sV-q}^{\mathrm{hc}}}]
	= \e^{-\beta (n-q)}\ \trace_{\mathcal{H}^{\mathrm{hc}}_{n-q,\sV-q,+}}
	\left[ \e^{\beta \mathcal{P}^{\mathrm{hc}}_{n-q}
		P^{n-q}_{\sV-q}\mathcal{P}^{\mathrm{hc}}_{n-q}} \right]
\end{equation}
while
\begin{align}\label{Z2}
	\ccPart
	&= \trace_{\mathcal{H}^{\mathrm{hc}}_{n-q,\sV-q,+}}
  		[\e^{-\beta H_{(\sV-q)/V,n-q,\sV-q}^{\mathrm{hc}}}]\nonumber
\\
  	&= \e^{-\beta (n-q)}\ \trace_{\mathcal{H}^{\mathrm{hc}}_{n-q,\sV-q,+}}
  		\left[ \e^{\beta(\frac{\sV-q}{\sV}) \mathcal{P}^{\mathrm{hc}}_{n-q} P^{n-q}_{\sV-q}
  		\mathcal{P}^{\mathrm{hc}}_{n-q}} \right].
\end{align}
Comparison of (\ref{Z1}) and (\ref{Z2}) yields
\begin{equation*}\label{}
     \ccPart=\e^{-\beta \frac{q}{V}(n-q)}\ Z_{\beta(\frac{\sV-q}{\sV})}(n-q,V-q)
\end{equation*}
and thus we have to analyse the following ratio:
\begin{equation}                            \label{c_expression} 
    \frac{ \e^{-\beta \frac{q}{V}(n-q)} Z_{\beta(\frac{\sV-q}{\sV})}(n-q, V-q)}{\cPart}.
\end{equation}
Penrose in \cite{Penrose} gave an explicit expression for $\cPart$:
\begin{equation*}\label{}
    \cPart=\hskip -0.3cm\sum_{r=0}^{\min(n, V-n)} z(r,n,V,\beta),
\end{equation*}
where
\begin{equation*}\label{}
    z(r,n,V,\beta):=\left( \frac{V-2r+1}{V-r+1} \right) \binom{V}{r}
    \exp\left\{ -\frac{\beta}{V} \left[ Vr - r^2 + r + n^2 - n \right] \right\}.
\end{equation*}
He also proved that if $h_\sV:[0,\min(\rho,1-\rho)]\to \RR$ converges uniformly in $[0,\min(\rho,1-\rho)]$
as $V\to\infty$ to a continuous  function $h:[0,\min(\rho,1-\rho)]\to \RR$, then
\begin{equation}\label{LD}
    \thermlim \frac{1}{\cPart} \sum_{r=0}^{\min(n,V-n)} h_\sV(\tfrac{r}{V})\
    z(r,n,V,\beta) =
    \begin{cases}
    h(\rho),
    & \mathrm{if}\ \  \rho\in[0,\rho_\beta],\\
    h(\rho_\beta),
    & \mathrm{if}\ \ \rho\in[\rho_\beta,1-\rho_\beta],\\
    h(1-\rho),
    & \mathrm{if}\ \  \rho\in[1-\rho_\beta,1].
    \end{cases}
\end{equation}
\par
We wish to express the ratio in (\ref{c_expression}) in the form of the lefthand side of
(\ref{LD}). We have
\begin{align*}
    Z_{\beta(\frac{\sV-q}{\sV})}(n-q, V-q)
    = \sum_{r=0}^{\min(n-q, V-n)} & \left( \frac{V-q-2r+1}{V-q-r+1} \right) \binom{V-q}{r} \\
    &   \times \exp\left\{ -\frac{\beta}{V} \left[ r(V-q) - r^2 + r + (n-q)^2 - (n-q) \right] \right\}
\end{align*}
For the case $\rho > \tfrac{1}{2}$, for large $V$, $n-q> V-n$ we must sum from
zero to $V-n$ and a straightforward calculation then gives
\begin{equation*}\label{}
    \e^{-\beta \tfrac{q}{V}(n-q)}Z_{\beta(\frac{\sV-q}{\sV})}(n-q, V-q)=\sum_{r=0}^{V-n} h_\sV(\tfrac{r}{V})\
    z(r,n,V,\beta)
\end{equation*}
where
\begin{multline} \label{h-rho>1/2}
    h_\sV(x) =  \left( \frac{1-2x-(q-1)/V}{1-2x+1/V} \right)
    \left( \frac{1-x+1/V}{1-x-(q-1)/V} \right) \\
        \times \prod_{s=0}^{q-1}\left(\frac{1-x-s/V}{1-s/V}\right)
        \exp\left\{ \beta q \left[ x + \rho - 1/V \right] \right\}.
\end{multline}
Therefore
\begin{equation} \label{h-lim>1/2}
    h(x)=\lim_{V \to\infty} h_\sV(x)=(1-x)^q \ \e^{q\beta (x+\rho)}.
\end{equation}
It is clear that the convergence is uniform since $h_\sV(x)$ is a product of terms each of which
converges uniformly on $[0,1-\rho]$ for $\rho>\tfrac{1}{2}$. Thus
\begin{equation*}
\thermlim \frac{ \e^{-\beta \frac{q}{V}(n-q)} Z_{\beta(\frac{\sV-q}{\sV})}(n-q, V-q)}{\cPart}
=
\begin{cases}
   (1-\rho_\sbeta)^q\ \e^{q\beta (\rho_\sbeta+\rho)}
    & \mathrm{if}\ \ \rho\in(1/2,1-\rho_\sbeta], \vspace{0.1cm}\\
    \rho^q \e^{\beta q}
    & \mathrm{if}\ \  \rho\in(1-\rho_\sbeta,1] .
    \end{cases}
\end{equation*}
Note that using the relation
\begin{equation*}
    \beta=\frac{1}{1-2\rho_\sbeta}\ln\left(\frac{1-\rho_\sbeta}{\rho_\sbeta}\right )
\end{equation*}
we get
\begin{equation*}
    (1-\rho_\sbeta)^q\ \e^{q\beta (\rho_\sbeta+\rho)}
    =\rho_\sbeta^q\ \e^{q\beta (1+\rho-\rho_\sbeta)}
\end{equation*}
and therefore we have proved Proposition \ref{Z} for $\rho>\tfrac{1}{2}$.
For the case $\rho \le \tfrac{1}{2}$ we have that $n-q < V-n$, the sum
for $\e^{-\beta \frac{q}{V}(n-q)} Z_{\beta(\frac{\sV-q}{\sV})}(n-q, V-q)$ is up to $n-q$,
and therefore we need to shift the index by $q$ to get it into the required form.
After shifting we get
\begin{align*}
\e^{-\beta \tfrac{q}{V}(n-q)} Z_{\beta(\frac{\sV-q}{\sV})}(n-q, V-q)
 &= \sum_{r=q}^{n} z(r,n,V,\beta) \left( \frac{V+q-2r+1}{V-2r+1} \right)\\
    & \hspace{0cm} \times \frac{ r(r-1)(r-2) \cdots (r-q+1)}{V(V-1)(V-2)\cdots (V-q+1)}
    \exp\left\{ \frac{\beta q}{V} \left[ V + n - r \right] \right\}.
\end{align*}
Note that summand is zero if we put $r=0, \dots, q-1$. Thus we may sum
from zero to $n$ to get as before
\begin{equation*}
    \e^{-\beta \tfrac{q}{V}(n-q)}Z_{\beta(\frac{\sV-q}{\sV})}(n-q, V-q)=\sum_{r=0}^{n} h_\sV(\tfrac{r}{V})\
    z(r,n,V,\beta)
\end{equation*}
where this time
\begin{equation} \label{h-rho<=1/2}
    h_\sV(x) =  \left( \frac{1-2x +(q+1)/V}{1-2x+1/V} \right)
     \prod_{s=0}^{q-1}\left( \frac{x -s/V}{1-s/V} \right)
                        \exp\left\{ \beta q \left[ 1 + \rho - x  \right] \right\}
\end{equation}
so that
\begin{equation*}
   h(x)=\thermlim h_\sV(x) = x^q \exp\{\beta q(1 + \rho - x) \}.
\end{equation*}
Convergence is again uniform on $[0,\rho]$ for $\rho<\tfrac{1}{2}$
and therefore
\begin{equation*}\label{}
\thermlim \frac{ \e^{-\beta \frac{q}{V}(n-q)} Z_{\beta(\frac{\sV-q}{\sV})}(n-q, V-q)}{\cPart}
=
\begin{cases}
       \rho^q \e^{\beta q}
    & \mathrm{if}\ \  \rho\in[0,\rho_\sbeta),  \vspace{0.1cm}\\
    \rho_\sbeta^q\ \e^{q\beta (1+\rho-\rho_\sbeta)}
    & \mathrm{if}\ \ \rho\in[\rho_\sbeta,1/2),
    \end{cases}
\end{equation*}
proving Proposition \ref{Z} for $\rho < \tfrac{1}{2}$. The case $\rho = \tfrac{1}{2}$
is more delicate because the first term in (\ref{h-rho<=1/2}) does not converge uniformly.
We can write (taking $V=2n$)
\begin{equation*}\label{}
    h_{2n}(r/2n)=\widetilde{h}_{2n}(r/2n)+\frac{q}{2(n-r)+1}\widetilde{h}_{2n}(r/2n)
\end{equation*}
where
\begin{equation*} \label{h-rho=1/2}
    \widetilde{h}_{2n}(x) =
     \prod_{s=0}^{q-1}\left( \frac{x -s/{2n}}{1-s/{2n}} \right)
                        \exp\left\{ \beta q \left[ 3/2 - x  \right] \right\}.
\end{equation*}
Clearly $\widetilde{h}_{2n}(x)$ converges uniformly on $[0,1/2]$ and therefore
\begin{equation*}
    \lim_{n\to\infty}\frac{1}{Z_\sbeta(n,2n)}\sum_{r=0}^{n} \widetilde{h}_{2n}(\tfrac{r}{2n})\
    z(r,n,2n,\beta)=\rho^q_\sbeta \, \mathrm{e}^{\beta q(3/2-\rho_\sbeta)}.
\end{equation*}
We thus have to show that
\begin{equation*}
    \lim_{n\to\infty}\frac{1}{Z_\sbeta(n,2n)}
    \sum_{r=0}^{n} \frac{\widetilde{h}_{2n}(\tfrac{r}{2n})}{2(n-r)+1}\
    z(r,n,2n,\beta)=0.
\end{equation*}
Since $\widetilde{h}_{2n}(x)$ is bounded, by $C$ say,
\begin{equation*}
    \lim_{n\to\infty}\frac{1}{Z_\sbeta(n,2n)}
    \sum_{r<n-n^{1/4}} \frac{\widetilde{h}_{2n}(\tfrac{r}{2n})}{2(n-r)+1}\
    z(r,n,2n,\beta)\leq \lim_{n\to\infty}\frac{C}{2n^{1/4}}=0.
\end{equation*}
On the other hand one can prove that for $n-2n^{1/2}\leq r\leq n-n^{1/2}$ and $r'\geq n-n^{1/4}$
\begin{equation*}\label{}
    \ln z(r,n,2n,\beta)-\ln z(r',n,2n,\beta)>\frac{1}{8}\ln n
\end{equation*}
for $n$ large, so that $z(r',n,2n,\beta)/z(r,n,2n,\beta)<1$.
Therefore
\begin{eqnarray*}
  \lim_{n\to\infty}\frac{1}{Z_\sbeta(n,2n)}
    &&\hskip -1cm\sum_{r\geq n-n^{1/4}} \frac{\widetilde{h}_{2n}(\tfrac{r}{2n})}{2(n-r)+1}\
    z(r,n,2n,\beta)
    \leq
    \lim_{n\to\infty}
    C\frac{\displaystyle{\sum_{r\geq n-n^{1/4}}\hskip -0.4 cm z(r,n,2n,\beta)}}
    {\displaystyle{\sum_{n-2n^{1/2}\leq r\leq n-n^{1/2}}\hskip -1.1cm z(r,n,2n,\beta)}} \\
   &\leq &  \lim_{n\to\infty}\frac{C}{n^{1/4}}
    \frac{\displaystyle{\max_{r\geq n-n^{1/4}}\hskip -0.3 cm z(r,n,2n,\beta)}}
    {\displaystyle{\min_{n-2n^{1/2}\leq r\leq n-n^{1/2}}\hskip -1.0cm z(r,n,2n,\beta)}}
\leq\lim_{n\to\infty}\frac{C}{n^{1/4}}=0.
\end{eqnarray*}

\subsection{Proof of Proposition \ref{c-q}}\label{sub c-q}
Recall that
\begin{align*}
\widetilde{c}\,^n_\sV(q)
&   = \frac{1}{Z_\sbeta(n,V)} \frac{1}{V} \trace_{\Hhccyclespace}
        \left[ U_q \e^{-\beta \widetilde{H}^{\mathsf{hc}}_{q,n,\sV} } \right].
\end{align*}
Considering the trace over $\Hhccyclespace$, expanding it in
terms of its basis $\{|\ii; \kk\ra\}$ and using Remark \ref{rem a} above,
where $\ii \sim \kk$
\begin{align*}
 \trace_{\Hhccyclespace}
        \left[ U_q \e^{-\beta \widetilde{H}^{\mathsf{hc}}_{q,n,\sV} } \right]
&   = \sum_{\kk} \sum_{\ii \sim \kk}
\la \ii; \kk | U_q \e^{-\beta \Phc (n - P_\sV^{(n-q)} ) \Phc}
        | \ii; \kk \ra\\
&   = \e^{-\beta q} \sum_{\kk} \sum_{\ii \sim \kk} \la U_q \ii; \kk | \e^{-\beta H^{\ii}}
        | \ii; \kk \ra
= \e^{-\beta q} \sum_{\kk} \sum_{\ii \sim \kk} \la U_q \ii | \ii \ra
        \la \kk | \e^{-\beta H^{\ii}} | \kk \ra.
\end{align*}
For $q > 1$, an element of the basis of the unsymmetrised $q$-space $\Hcan{q}$ may
be written as an ordered $q$-tuple $\ii = (i_1, i_2, \dots , i_q)$ where the $i_l$'s are
all distinct.
Then we may write
\begin{align*}
\la U_q \ii | \ii \ra
&   = \la U_q (\mathbf{e}_{i_1} \otimes \mathbf{e}_{i_2} \otimes \dots \otimes \mathbf{e}_{i_q} ) \, |\,
        \mathbf{e}_{i_1} \otimes \mathbf{e}_{i_2} \otimes \dots \otimes \mathbf{e}_{i_q} \ra
\\
&   = \la \mathbf{e}_{i_2} \otimes \mathbf{e}_{i_3} \otimes \dots \otimes \mathbf{e}_{i_q} \otimes \mathbf{e}_{i_1}\, |\,
        \mathbf{e}_{i_1} \otimes \mathbf{e}_{i_2} \otimes \dots \otimes \mathbf{e}_{i_q}\ra=0.
\end{align*}
Hence $\widetilde{c}\,_\sV^n(q)$ is non-zero only if $q=1$.

For the second statement, note that we may re-express $\widetilde{c}\,_\sV^n(1)$ as follows:
\begin{align*}
    \widetilde{c}\,_\sV^n(1)
&   = \frac{1}{Z_\sbeta(n,V)} \frac{1}{V} \trace_{\Phc(\Hcan{1} \otimes \Hcansym{n-1})}
        \left[ \e^{-\beta \widetilde{H}^{\mathsf{hc}}_{1,n,\sV} } \right]
\\
&   = \frac{\e^{-\beta}}{Z_\sbeta(n,V)} \frac{1}{V} \sum_{i=1}^V \sum_{\kk\, /\hskip-0.17cm \ni\, i}
        \la \kk | \e^{-\beta H^i } | \kk \ra
\\
&   = \frac{\e^{-\beta}}{Z_\sbeta(n,V)} \frac{1}{V} \sum_{i=1}^V \trace_{\Hhccansym{n-1}}
        \left[ \mathcal{P}_i \e^{-\beta H^i} \mathcal{P}_i \right]
\\
&   = \e^{-\beta} \, \frac{ Z_\sbeta(\tfrac{V-1}{V}, n-1, V-1, )}{ Z_\sbeta(n,V) }
\end{align*}
and the result follows from Proposition \ref{Z}.
\section{ODLRO}
The one-body reduced density matrix for $x,x'\in \Lambda_\sV $ may be defined as
\begin{equation}\label{}
	D_{\sbeta,n,\sV}(x,x'):=\la a^*_x a\starr_{x'} \ra=\frac{1}{\cPart}
	\trace_{\Hhccansym{n}} \left[ K^{(n)}_{x,\,x'} \e^{-\beta \hcHam} \right].
\end{equation}
where for $\phi\in \mathcal{H}_\sV$, $K_{x,\,x'}\phi=\la \mathbf{e}_{x'}|\phi \ra \mathbf{e}_x$.

Penrose showed that for $x\neq x'$,
\begin{equation*}
	\thermlim D_{\sbeta,n,\sV}(x,x')=\rho_\sbeta^c ,
\end{equation*}
that is, whenever Bose-Einstein condensation occurs, there is
\textit{Off-diagonal long-range order} as defined by Yang \cite{Yang}.
It has been argued and proved in some cases (see for example \cite{Ueltschi2} and \cite{DMP})
that in the expansion of $D_{\sbeta,n,\sV}(x,x')$ in terms of permutation cycles, only infinite cycles contribute to
long-range order. Here we are able to show this explicitly.
\par
By the proposition in Appendix \ref{appendixA}, we have
\begin{equation*}
	D_{\sbeta,n,\sV}(x,x')= \sum_{q=1}^n C_\sV^n(q;K_{x,\,x'})
\end{equation*}
where
\begin{equation}
	C^n_\sV(q;K_{x,\,x'})
	= \frac{1}{Z_\sbeta(n,V)} \, \trace_{\Hhccyclespace}
	\left[ (K_{x,\,x'}\otimes I\otimes I\otimes \ldots\otimes I)
 	U_q \e^{-\beta \hcHam } \right].
\end{equation}
Note that this is equivalent to the expansion of $\sigma_\rho(x)$ in equations (2.14) and (2.16) in \cite{Ueltschi2}.

Applying the argument in Subsections \ref{sub c} and \ref{sub Z}, we can show that
\begin{equation} \label{proj_in_qspace_irrel}
	\thermlim C^n_\sV(q;K_{x,\,x'})=\thermlim \widetilde{C}^n_\sV(q;K_{x,\,x'})
\end{equation}
where we take
\begin{equation*}
	\widetilde{C}^n_\sV(q;K_{x,\,x'}) = \frac{1}{Z_\sbeta(n,V)} \, \trace_{ \Hhccyclespace  }
	\left[ (K_{x,\,x'}\otimes I\otimes I\otimes \ldots\otimes I)
 	U_q \e^{-\beta \widetilde{H}^{\mathsf{hc}}_{q,n,\sV} } \right].
\end{equation*}
The only difference is that instead of equation (\ref{sum_of_cycled_inner_products}), we obtain
\begin{multline} \label{summs}
	\qquad \sum_{\ii^0} \sum_{i^1_{r_1} \notin \ii^0 \setminus \{i^0_{r_1}\}} \hspace{0.1cm}
	\sum_{i^2_{r_2} \notin [r_1, i^1_{r_1}](\ii^0) \setminus \{i^1_{r_2}\}} \!\!
	\cdots
	 \!\! \sum_{i^m_{r_m} \notin [r_{m-1}, i^{m-1}_{r_{m-1}}; \, \dots \, ; r_1, i^1_{r_1}](\ii^0)
	\setminus \{i^{m-1}_{r_m}\}}
\\
	\la [r_m, i^m_{r_m}; \, \dots \, ; r_2, i^2_{r_2} ; r_1, i^1_{r_1}](\ii^0) |
		(K_{x,x'} \otimes I \otimes \cdots \otimes I)U_q \ii^0 \ra \qquad
\end{multline}
whose treatment is similar but slightly more complicated, as detailed below.

Let $q > 1$ and consider the case $\{r_1, r_2, \dots , r_m\} \neq \{1,2,\dots, q\} $.
When $1 \notin \{r_1, r_2, \dots , r_m\}$ we obtain inner products of the form:
\[
	\la i^0_1 | K_{x,x'} i^0_2 \ra \la j_2, j_3, \dots , j_q | i^0_3, i^0_4, \dots , i^0_q, i^0_1 \ra
\]
where $j_k \ne i^0_1$ for all $k$ by the hard-core condition, implying the second term is zero
as $j_q \ne i^0_1$. On the other hand, when $1 \in \{r_1, r_2, \dots , r_m\}$, then there exists at least one
$l \notin \{r_1, r_2, \dots , r_m\}$, yielding an inner product of the form
\[
	\la j_1 | K_{x,x'} i^0_2 \ra \la j_2, \dots , j_{l-1}, i_l, j_{l+1}, \dots, j_q | i^0_3, i^0_4, \dots , i^0_q, i^0_1 \ra
\]
which also results in the second term being zero as $\la i_l | i_{l+1} \ra = 0$. Note that the above cases do
not occur for $q=1$.

For the case $\{r_1, r_2, \dots , r_m\} = \{1,\dots, q\} $, as before, the remaining sum may be bounded by a similar
expression whose summations have slightly relaxed restrictions. Also the left hand side of the inner product is
independent of $\ii^0$, so again denoting $\mathbf{s}^0 = (1,2,3,\dots,q)$, we have
\begin{align*}
(\ref{summs})&
	\le
	\underbrace{ \sum_{i^1_{r_1}=1}^\sV \;\;
	\sum_{i^2_{r_2}=1}^\sV \;\;
	\cdots  \;\;
	\sum_{i^m_{r_m} = 1}^\sV }_{\stackrel{\text{where $[r_m, i^m_{r_m}; \, \dots
		\, ; r_2, i^2_{r_2} ; r_1, i^1_{r_1}](\mathbf{s}^0)$}}{\text{\tiny{has distinct indices}}}} \!\!
	\sum_{\ii^0} \;
	\la[r_m, i^m_{r_m}; \, \dots \, ; r_2, i^2_{r_2} ; r_1, i^1_{r_1}]
		(\mathbf{s}^0) | (K_{x,x'} i^0_2), i^0_3, \dots , i^0_q, i^0_1 \ra
\intertext{and as there is only one possible value for each $i^0_1, i^0_3, i^0_4, \dots , i^0_q$
giving a non-zero summand, we can bound above by}
&\le 	\sum_{i^1_{r_1}=1}^\sV \;
	\sum_{i^2_{r_2}=1}^\sV \;
	\cdots  \;
	\sum_{i^m_{r_m} = 1}^\sV
	\sum_{i^0_2=1}^\sV \la i^k_{r_k} | K_{x,x'} i^0_2 \ra \,
=	\, V^{m-1} \sum_{i^k_{r_k} = 1}^\sV \sum_{i^0_2=1}^\sV \la i^k_{r_k} | K_{x,x'} i^0_2 \ra \,
=	\,V^{m-1}
\end{align*}
where $k \in [1,m]$ is the smallest number such that $r_k = 1$, and for any $x, x' \in \Lambda_\sV$. Thus
the entire sum (\ref{summs}) is bounded above by $V^{m-1}$. Therefore one can conclude the argument
of Subsection \ref{sub Z}, proving (\ref{proj_in_qspace_irrel}).

Moreover, following the reasoning in Subsection \ref{sub c-q},
we can then check that for $q \ge 1$ and $x\neq x'$,
$\widetilde{C}^n_\sV(q;K_{x,\,x'})=0$, since for $q=1$, $\la \mathbf{e}_i | K_{x,x'} \mathbf{e}_i \ra = 0$, and for $q > 1$
\[
\la (K_{x,x'} \otimes I \otimes \dots \otimes I) U_q \ii | \ii \ra
    = \la (K_{x,x'} \mathbf{e}_{i_2}) \otimes \mathbf{e}_{i_3} \otimes \dots
    	\otimes \mathbf{e}_{i_q} \otimes \mathbf{e}_{i_1}\, |\,
	\mathbf{e}_{i_1} \otimes \mathbf{e}_{i_2} \otimes \dots \otimes \mathbf{e}_{i_q}\ra=0
\]
as the $i_l$'s are all distinct. So we have that
\begin{equation*}
	\thermlim C^n_\sV(q;K_{x,\,x'})=0
\end{equation*}
and that
\begin{equation*}
	\lim_{Q\to \infty} \thermlim \sum_{q=Q+1}^\infty C^n_\sV(q;K_{x,\,x'})=
	\thermlim D_{\sbeta,n,\sV}(x,x')=\rho_\sbeta^c.
\end{equation*}

\appendix
\renewcommand{\theequation}{A.\arabic{equation}}
\setcounter{equation}{0}  
\section{Appendix: Expectation of Operator in Terms of Cycle Lengths}  \label{appendixA}
\begin{proposition}
	Given an operator $A$ on $\Hone$, the expectation of $A$ may be expressed in terms
	of cycle lengths
	\begin{equation}
		\langle A^{(n)} \rangle = \sum_{q=0}^n C^n_\sV(q;A)
	\end{equation}
	where
	\begin{equation}
		C^n_\sV(q;A)
		= \frac{1}{Z_\sbeta(n,V)} \, \trace_{\Hhccyclespace}
		\left[ (A\otimes I\otimes \ldots\otimes I)
	 	U_q \e^{-\beta \hcHam } \right].
	\end{equation}
	\end{proposition}
Note that $c^n_\sV(q)=C^n_\sV(q;I)/V$ where $c^n_\sV(q)$ is as defined in Proposition \ref{stat}.

\begin{proof}
\begin{equation*}
	\langle A^{(n)} \rangle
	= \frac{1}{\cPart}
				\trace_{\Hhccansym{n}} \left[ A^{(n)} \e^{-\beta \hcHam} \right] \\
	= \frac{1}{\cPart} \frac{1}{n!} \sum_{\pi \in S_n}
				\trace_{\Hhccan{n}} \left[ A^{(n)} U_\pi \e^{-\beta \hcHam} \right]
\end{equation*}
using the facts that $[U_\pi,\hcHam]=0$, $[U_\pi,\Phc]=0$ and by the cyclicity of the trace. Note we can simplify this
expression by the following method:
\begin{align*}
	\trace_{\Hhccan{n}} \left[ A^{(n)} U_\pi \e^{-\beta \hcHam} \right]
		&= 	\trace_{\Hhccan{n}} \left[ \sum_{i=1}^n
				(I \otimes \cdots \otimes \underbrace{A}_{\text{ $i$th position}} \otimes \cdots \otimes I)
				U_\pi \e^{-\beta \hcHam} \right] \\
		&= 	\trace_{\Hhccan{n}} \left[ \sum_{i=1}^n U_{(1i)} (A \otimes I \otimes \cdots \otimes I)
				U_{(1i)} U_\pi \e^{-\beta \hcHam}\right]
\intertext{\vspace{-0.4cm}where $U_{(1i)}$ represents the transposition $(1 \, i)$, so using cyclicity of the trace again}
		&= 	\trace_{\Hhccan{n}} \left[ \sum_{i=1}^n (A \otimes I \otimes \cdots \otimes I)
				U_{(1i)} U_\pi \e^{-\beta \hcHam} U_{(1i)} \right] \\
		&= 	\trace_{\Hhccan{n}} \left[ \sum_{i=1}^n (A \otimes I \otimes \cdots \otimes I)
				U_{(1i)} U_\pi U_{(1i)} \e^{-\beta \hcHam} \right] \\
		&= 	n \; \trace_{\Hhccan{n}} \left[ (A \otimes I \otimes \cdots \otimes I) U_{\pi'}  \e^{-\beta \hcHam} \right]
\end{align*}
where $\pi' = (1\,i) \, \pi \, (i\,1)$ using (\ref{invar_under_cycles}). Thus
\begin{equation}
	\langle A^{(n)} \rangle
	=  \frac{1}{\cPart} \frac{1}{(n-1)!} \sum_{\pi \in S_n}     \label{sum_perms}            
		\trace_{\Hhccan{n}} \Big[ (A \otimes I \otimes \cdots \otimes I) U_\pi \e^{-\beta \hcHam} \Big] .
\end{equation}

Given distinct indices $i_2, \dots, i_q$, let
\[
	S_n^q(i_2, i_3, \dots i_q)
	= \Big\{ \pi \in  S_n : \pi(i_m) = i_{m+1},  1 \le m < q \text{ with }  i_1 = \pi(i_q) = 1 \Big\}.
\]
Then for any $\pi \in S_n^q(i_2, i_3, \dots i_q)$, there exists a
$\pi' \in S_{n-q}$ so that one can write
\[
	\trace_{\Hhccan{n}} \Big[ ( A \otimes I \otimes \cdots \otimes I ) U_\pi \e^{-\beta H} \Big]
	= \trace_{\Hhccan{n}} \Big[ ( A \otimes I \otimes \cdots \otimes I )(U_q \otimes U_{\pi'}) \e^{-\beta H} \Big].
\]
The set $S_n^q(i_2, i_3, \dots i_q)$ form a partition of the set of permutations where $1$ belongs
to a cycle of length $q$. There are $\tfrac{(n-1)!}{(n-q)!}$ such sets. Then
\begin{align*}
	\langle A^{(n)} \rangle
	&=  \frac{1}{\cPart} \frac{1}{(n-1)!} \sum_{\pi \in S_n}
			\trace_{\Hhccan{n}} \Big[ (A \otimes I \otimes \cdots \otimes I) U_\pi \e^{-\beta \hcHam} \Big]
\\
	&= \frac{1}{\cPart} \frac{1}{(n-1)!} \sum_{q=1}^n \frac{(n-1)!}{(n-q)!}
			\sum_{\pi' \in S_{n-q}}
			\trace_{\Hhccan{n}} \Big[ (A \otimes I \otimes \cdots \otimes I)
				(U_q \otimes U_{\pi'})  \e^{-\beta \hcHam} \Big]
\\
	&= \frac{1}{\cPart} \sum_{q=1}^n \frac{1}{(n-q)!}  \sum_{\pi' \in S_{n-q}}
			\trace_{\Hhccan{n}} \Big[ (A \otimes I \otimes \cdots \otimes I)
				(U_q \otimes U_{\pi'})  \e^{-\beta \hcHam} \Big]
\\
	&= \frac{1}{\cPart} \sum_{q=1}^n \trace_{\Phc (\Hcan{q} \otimes \Hcansym{n-q})}
			\Big[ (A \otimes I \otimes \cdots \otimes I) U_q  \e^{-\beta \hcHam} \Big]
\end{align*}
and recall that $\Hhccyclespace := \Phc ( \Hcan{q} \otimes \Hcansym{n-q} )$.
\end{proof}

\textbf{Acknowledgements:} The authors would like to thank T.C. Dorlas and S. Adams for some helpful discussions and 
they are grateful to the referees for their many useful suggestions. GB would like to thank the Irish Research Council for
Science, Engineering and Technology for their financial support.  

%

\end{document}